\documentclass[a4paper,11pt]{article}
\pdfoutput=1
\usepackage{jheppub}
\usepackage[utf8]{inputenc}
\usepackage{epsfig,amsfonts,amsthm}
\usepackage{amsmath,amssymb}
\usepackage{float}
\usepackage{xcolor}
\usepackage{braket}
\usepackage{booktabs}

\newcommand{\be}{\begin{equation}}
\newcommand{\ee}{\end{equation}}
\newcommand{\bea}{\begin{eqnarray}}
\newcommand{\eea}{\end{eqnarray}}

\newcommand{\Z}{\mathbb{Z}}
\newcommand{\mmatrix}[4]{ \left(\! \begin{array}{ccc}#1 & #2 \\ #3 & #4 \end{array}\!\right) }
\providecommand{\id}{{\boldsymbol{1}}}
\newcommand{\diag}{\mathrm{diag}}
\newcommand{\Aut}{\mathrm{Aut}}
\newcommand{\fdf}[2]{(\phi^\dagger_{#1}\phi_{#2})}
\newcommand{\fdfn}[2]{\phi^\dagger_{#1}\phi_{#2}}

\definecolor{darkred}{rgb}{0.7,0.0,0.0}

\newcommand{\gray}{\color{gray}}
\definecolor{darkgreen}{rgb}{0.0,0.5,0.0}

\def\lsim{\mathrel{\rlap{\lower4pt\hbox{\hskip1pt$\sim$}}
    \raise1pt\hbox{$<$}}}         
\def\gsim{\mathrel{\rlap{\lower4pt\hbox{\hskip1pt$\sim$}}
    \raise1pt\hbox{$>$}}}         

\title{Symmetries for the 4HDM: extensions of cyclic groups}

\author[1]{Jiazhen Shao,}
\author[2]{Igor P. Ivanov}
\affiliation{School of Physics and Astronomy, Sun Yat-sen University, 519082 Zhuhai, China}
\emailAdd{shaojzh5@mail2.sysu.edu.cn} 
\emailAdd{ivanov@mail.sysu.edu.cn} 

\abstract{
Multi-Higgs models equipped with global symmetry groups, either exact or softly broken,
offer a rich framework for constructions beyond the Standard Model and 
lead to remarkable phenomenological consequences.
Knowing all the symmetry options within each class of models can guide its phenomenological exploration,
as confirmed by the vast literature on the two- and three-Higgs-doublet models.
Here, we begin a systematic study of finite non-abelian symmetry groups
which can be imposed on the scalar sector of the four-Higgs-doublet model (4HDM) without leading to accidental symmetries.
In this work, we derive the full list of such non-abelian groups available in the 4HDM 
that can be constructed as extensions of cyclic groups by their automorphism groups.
This list is remarkably restricted but it contains cases which have not been previously studied.
Since the methods we develop may prove useful for other classes of models,
we present them in a pedagogical manner.
}

\begin{document}
\maketitle


\section{Introduction}

\subsection{Beyond three Higgs doublets: the early history and recent works}

One may doubt that time is ripe for studying the four-Higgs-doublet models (4HDM), 
given that only one Higgs boson has been found in experiment so far \cite{CMS:2012qbp,ATLAS:2012yve}.
However, the logic of scientific inquiry and the recent proliferation of multi-scalar models
make this objective valid and timely.
In fact, investigation of the opportunities offered by the 4HDMs started off more than 40 years ago, 
soon after T.~D.~Lee suggested the two-Higgs-doublet model (2HDM) to break the $CP$ symmetry spontaneously \cite{Lee:1973iz}
and S.~Weinberg found a way to combine $CP$ violation with natural flavor conservation 
within the three-Higgs-doublet model (3HDM) \cite{Weinberg:1976hu}
(for recent reviews, see \cite{Branco:2011iw,Ivanov:2017dad};
an introduction to finite symmetry groups used in particle physics can be found in \cite{Ishimori:2010au}).
In the past decades, the history of the 4HDMs even went through several episodes of a rather intense activity,
and we feel it appropriate to give a short historical summary of this research.

To our knowledge, the first publication which explicitly used four Higgs doublets 
was the 1977 paper by Bjorken and Weinberg \cite{Bjorken:1977vt}.
The authors discussed the idea of possible muon number non-conservation arising through scalar fields mixing.
Considering only two lepton generations, the general Yukawa structure in the electron-muon sector can be written,
in modern notation, as $\bar L_{L\, a} \Gamma^{i}_{ab} \phi_i \ell_{R\, b} + h.c.$,
where each $\Gamma^i$ is a complex $2\times 2$ matrix.
The idea of \cite{Bjorken:1977vt} was that each entry of this set of coupling matrices
is generated by its own Higgs doublet $\phi_i$, $i = 1,\dots, 4$.

Exploration of $N$-Higgs-doublet models (NHDM) exploded in 1979 after the $b$ quark 
was experimentally observed and the quark masses and mixing parameters hinted 
at a hierarchical structure, begging for explanation.
The tantalizing idea entertained by many was to actually deduce this structure
using group theory arguments within NHDMs equipped with global discrete symmetry groups.
Most work in this directions naturally dealt with the 3HDM, but occasional
remarks concerning 4HDM or beyond were also made.
In one of the earliest papers \cite{Wyler:1979fe}, Wyler explored the $S_4$-symmetric 3HDM
with the Higgses transforming as a triplet under $S_4$.
One of the conclusions was that the $S_4$-invariant potential retained residual symmetries
at any of its minima. However, by adding the fourth Higgs doublet transforming as a singlet under $S_4$,
spontaneous breaking of the full $S_4$ could be achieved.

It is instructive to mention that the question of whether a global flavor symmetry of an NHDM
is broken completely upon minimization is of extreme importance for achieving a viable quark and lepton sectors.
As first proved in \cite{Leurer:1992wg} and polished in \cite{GonzalezFelipe:2014mcf},
residual flavor symmetries lead to pathologies in the quark or lepton sectors, see
\cite{GonzalezFelipe:2013yhh,GonzalezFelipe:2013xok}
for explicit examples within the $A_4$-symmetric and $S_4$-symmetric 3HDMs.
This obstacle, however, is of a purely algebraic origin and could be avoided if a fourth doublet is added. 
Introducing soft breaking terms is also a way out \cite{Bree:2023ojl} 
but it comes at the expense of losing the predictive power of the fully symmetric models.
In short, the fourth Higgs doublet helps achieve what was impossible within a symmetry-constrained 3HDM.

Returning to the early days of the NHDM exploration, 
several models based on the permutation symmetry groups $S_n$, 
such as \cite{Pakvasa:1978tx,Ma:1979we,Derman:1979nf,Christos:1984if}, mentioned the 4HDM option.
Later into 1980's, when it became clear that no simple group-theoretic argument
could accommodate the quark parameters, including the uncomfortably large top quark mass,
other symmetry options were tried. For example, \cite{Rajpoot:1988gw}
proposed the 4HDM based on sign flips and discrete rephasing transformations.
As a curiosity, we mention the 1985 paper \cite{Kobayashi:1985ek} 
in which the 4HDM was also studied in the context of a preon-based model 
of composite leptons, quarks, and scalar bosons.

Exploration of models based on four Higgs doublets continued through the 1980's and 1990's,
with the focus either on the properties of the scalar bosons themselves \cite{Albright:1979yc,Drees:1988fc,Haber:1989xc,Griest:1989ew,Griest:1990vh}
or on the $CP$ violation in the scalar section \cite{Masip:1995sm}.
For an early systematic phenomenological study applicable to the generic NHDM, see \cite{Grossman:1994jb}.

In 1991, Ma proposed a 4HDM based on the symmetry group $S_3\times \Z_3$,
with the four Higgs doublets transforming as $1+1'+2$ under this group \cite{Ma:1990qh,Ma:1991eg}.
It offered a rather economic description of several quark and lepton sector regularities, 
including the tentative signal of the 17 keV neutrino \cite{Duong:1991yt}, 
a hot topic at that time \cite{Franklin:1995pk}.
The model has received some attention during 1990's, 
\cite{Deshpande:1991zh,Lavoura:1991yh,Lavoura:1992xj,Deshpande:1993py,Lavoura:1999dn}.

The recent activity on symmetry based explanations of the fermion properties within the 4HDMs
rely on various schemes to couple different Higgs doublets to different fermions.
The symmetry group is often the (softly broken) $\Z_2 \times \Z_2 \times \Z_2$,
but the exact assignment is a matter of choice.
In \cite{Cree:2011uy}, the authors begin with the so-called democratic 3HDM, in which 
the three doublets couple individually to the up-quarks, down-quarks and leptons, 
and then the fourth Higgs doublet is introduced to relax the constraints arising in the 3HDM case.
Although the fourth doublet decouples from the fermions, it offers more degrees of freedom for scalar mixing.
A similar idea, but in a supersymmetric setting, was used \cite{Arroyo-Urena:2019lzv}, 
where the fourth doublet was needed for anomaly cancellation. 
Another twist of this idea was proposed in \cite{Rodejohann:2019izm}
where the fermions were grouped into sets of similar masses: $\{t\}$, $\{b, c, \tau\}$, $\{\mu, s\}$, and $\{d, u, e\}$.
Each set was assumed to couple to its own Higgs doublet and, with all couplings of order one,
the fermion mass hierarchy could be explained by the hierarchy of vacuum expectation values (vevs).
In another, very recent, paper \cite{Goncalves:2023ydf}, one Higgs doublet was assumed to couple to all quarks,
while the other three doublets coupled to the electron, muon, and tau, respectively.
The ultimate realization of this idea is the ``Private Higgs'' framework of Porto and Zee
where each type of fermions receives its private Higgs doublet \cite{Porto:2007ed,Porto:2008hb},
but this construction brings us beyond four doublets.

Other recent 4HDMs addressing the fermion properties but based on different symmetry groups 
are \cite{CarcamoHernandez:2021osw}, making use of the four doublets organized into a $\Delta(27)$ triplet
and a singlet, and \cite{Vien:2020aif} based on the symmetry group $Q_6$. 

A global symmetry of an NHDM can remain unbroken at the minimum. 
If it acts trivially on fermions and if the entire lagrangian respects this symmetry, 
then it stabilizes one or more scalars against decay, rendering them the dark matter (DM) candidates.
The prototypical model is the $\Z_2$-symmetric 2HDM considered first
by Deshpande and Ma in 1977 \cite{Deshpande:1977rw}
and known today as the inert doublet model.
Starting from 2000's, many other multi-doublet realizations of this idea were proposed.
They include the 4HDM examples based on the groups $D_4\times \Z_n$ \cite{Meloni:2011cc,Lavoura:2011ry},
the group $A_4$ \cite{Meloni:2010sk,Boucenna:2011tj,deAdelhartToorop:2011ad,Bonilla:2023pna},
various cyclic groups \cite{Ivanov:2012hc},
or a general construction with $3+1$ Higgs doublets, with one of them being inert
\cite{Diaz-Cruz:2014pla}.
DM model-building with inert doublets went even beyond the 4HDM, see, for example, \cite{Keus:2014isa}.
We stress again that, if the unbroken global symmetry acts non-trivially on fermions,
the symmetry-constrained Yukawa sector will unavoidably generate pathologies in the quark and lepton masses or mixing,
as explained in \cite{Leurer:1992wg,GonzalezFelipe:2014mcf}.
The only way for an NHDM with unbroken symmetries to be phenomenologically viable
is to limit these symmetries to the inert Higgs doublets only and to decouple them from fermions.

Symmetry-based NHDMs are often used in neutrino mass models,
especially in those which aim to derive the tribimaximal mixing pattern
from group-theoretic relations.
In 2001, a very popular $A_4$-based 4HDM for neutrino masses and mixing was suggested by Ma and Rajasekaran \cite{Ma:2001dn}.
Other examples of neutrino models which make use of four Higgs doublets (and, possibly, additional scalars) 
and various symmetry groups can be found in \cite{He:2006dk,Grimus:2008tt,Grimus:2008nf,Grimus:2008vg,Grimus:2009sq,Grimus:2009pg,Ferreira:2011hw,Park:2011zt,BenTov:2012tg,Grossman:2014xia}.

Another path leading to the 4HDM is supersymmetrization of the usual (non-supersymmetric) 2HDM
\cite{Drees:1988fc,Griest:1989ew,Griest:1990vh,Nelson:1993vc,Krasnikov:1993qd,Masip:1995sm,Aranda:2000zf,Marshall:2010qi,Kawase:2011az,Clark:2011cv,Yagyu:2012qp,Grossman:2014xia,Dutta:2018yos}.
Explored since late 1980's, such models offer a setting which is less constrained than the minimal supersymmetric extension 
of the Standard Model.
Also, the 4HDMs in disguise arise in the so-called Twin Higgs models, 
where the visible sector and the twin sector, each, contain a pair of Higgs doublets
\cite{Chacko:2005pe,Chacko:2005vw,Falkowski:2006qq,Chang:2006ra,Yu:2016bku,Yu:2016swa,Yu:2016cdr}.
Although these are doublets under two different gauge groups, their scalar interaction
is equivalent to a class of 4HDMs.

Finally, in a broader context, there exist other models of neutrino mass generation or of sophisticated dark sectors 
which involve several, often very many, scalar fields. 
Details of the scalar sector in such models are often mentioned in passing, 
but they can contain symmetry-driven features or limitations.
Just to give a recent example of a less-studied construction, 
\cite{Belanger:2022esk} investigated the scenario with three dark matter candidates shaped by the symmetry group $\Z_7$.
We expect that the methods and the results which we present here for the 4HDM will be of relevance
to these multi-scalar models equipped with global symmetries.

\subsection{The quest for symmetries in the 4HDMs}

With the total number of papers dealing with four Higgs doublets 
approaching one hundred, it seems timely to bring some order to the plethora of symmetry-related 4HDMs.
It is true that many papers cited above involve not only the four Higgs doublets but also additional fields. 
To make the questions well-posed, let us focus on the 4HDMs proper, that is, on models
which contain four Higgs doublets with identical gauge quantum numbers and no other fields beyond the Standard Model.
One can now ask the following major questions:
\begin{itemize}
\item 
What is the full list of symmetry groups available in the 4HDM, 
either within the scalar sector alone or with fermions included?
\item
Are these symmetry groups implemented in the 4HDM uniquely, or do there exist inequivalent realizations?
\item
Can these symmetries be detected with basis-independent methods?
\item
How are these symmetry groups broken upon the potential minimization?
\item
Finally, what novel symmetry-driven phenomenological and astroparticle situations emerge within each of these symmetry settings?
\end{itemize}
Undoubtedly, this list represents a long research program, not a topic for a single paper.
But this program seems to be doable.
Indeed, within the 3HDM, many of these questions have already been answered,
including the classification of all finite symmetry groups in the scalar sector 
\cite{Ivanov:2011ae,Ivanov:2012fp,Ivanov:2012ry} and of
all abelian symmetry groups either in the scalar sector alone \cite{Ivanov:2011ae} or with quarks included \cite{Ivanov:2014doa},
insights into continuous symmetry groups of the scalar potential \cite{Darvishi:2019dbh,Darvishi:2021txa},
a new form of $CP$ symmetry of higher order \cite{Ivanov:2015mwl,Haber:2018iwr},
the list of all symmetry-breaking pattern for each finite group available \cite{Ivanov:2014doa},
and basis-independent detection of symmetry groups \cite{deMedeirosVarzielas:2019rrp}.

This comes together with already hundreds of theoretical and phenomenological papers on specific 3HDMs,
see \cite{Ivanov:2017dad} for an overview.
In many cases, the phenomenological works benefited from the systematic study of all symmetries available in the 3HDM. 
We expect that the role of systematic classification of symmetry-related aspects of the 4HDMs will be even higher
than for the three-doublet case due to the increased complexity and proliferation of free parameters.

\subsection{The goals of this work}

The goal of this paper is to give a partial answer to the first question in the above list.
Namely, we will find the full list of all discrete non-abelian symmetry groups in the 4HDM scalar sector
which can be constructed as extensions of cyclic groups without producing accidental symmetries.

To achieve this goal, we will need to further develop and apply the same group extension method
which was used in \cite{Ivanov:2012fp,Ivanov:2012ry} for the complete classification
of finite symmetry groups in the 3HDM scalar sector.
However, as we will show, some subtle group-theoretic arguments 
which worked for the 3HDM case no longer apply to the 4HDMs.
As a result, we will not be able to exhaust all finite non-abelian groups in the 4HDM using this procedure.
Still, we will get a class of symmetry-based 4HDMs, some of which have never been studied or mentioned before.
Moreover, some of them represent novel opportunities which do not exist in the 3HDM.
These novelties include not only the choice of the symmetry groups but also the way 
they are implemented in the NHDM setting.

We expect that at least some of the non-abelian symmetry groups constructed in this way 
will lead to 4HDMs with peculiar phenomenological features.
Such a phenomenological study should be the subject of a dedicated study.
In this work, we focus only on establishing the full list of non-abelian extensions of cyclic groups
which can lead to novel 4HDMs.

The structure of the paper is the following.
In the next Section, we recapitulate the group theoretic methods within the 3HDM,
focusing on those which lead to the 3HDM discrete group classification of \cite{Ivanov:2012fp,Ivanov:2012ry}.
Then in Section~\ref{section-4HDM-general}, we outline the main strategy for the 4HDM, 
give the list of available abelian group and their automorphism groups, 
and discuss the generic patterns
which we will later find in specific cases.
Section~\ref{section-4HDM-cases} contains the main results of the paper:
we go through the full list of cyclic groups $A$ available in the 4HDM and try to extend them
by automorphisms in all self-consistent ways.
The overall picture which emerges from this study is presented in Section~\ref{section-conclusions}.
Several Appendices provide auxiliary technical details
and introduces the Python code {\tt 4HDM Toolbox} which we used to make sure that no cases are missed.

\section{Discrete symmetry groups for the 3HDM scalar sector}\label{section-3HDM-recap}

\subsection{General remarks}

	In the NHDM, we introduce $N$ Higgs doublets $\phi_i$, $i = 1, \dots, N$, 
all possessing the same gauge quantum numbers, and
construct the Higgs lagrangian, which includes the scalar-gauge interactions 
coming from the kinetic terms and the renormalizable scalar self-interaction potential $V$.
The potential can be generically written as 
\begin{equation}
V = m_{ij}^2 \fdf{i}{j} + \lambda_{ijkl}\fdf{i}{j}\fdf{k}{l}\,,\label{V-NHDM}
\end{equation}
with indices going from 1 to $N$. In this generic expression,
the parameters $m_{ij}^2$ and $\lambda_{ijkl}$ satisfy the relations which guarantee that 
the potential is hermitian, such as $m_{ji}^2 = (m_{ij}^2)^*$ and so on.
Finally, the NHDM lagrangian also contains the Yukawa sector 
which describes the coupling of the Higgs doublets to fermions.
In this work, however, we will focus on the scalar potential.

It may happen that the Higgs lagrangian is invariant under a set of global transformations
$\phi_i \mapsto U_{ij}\phi_j$, where $U \in U(N)$.
Such a transformation represents a family symmetry of a given Higgs lagrangian;
the set of all its family symmetries forms the symmetry group $\tilde G \subseteq U(N)$. 

The group $U(N)$ contains the $U(1)$ subgroup of the common phase rotations of all doublets by the same phase shift. 
By construction, any Higgs lagrangian is automatically invariant under this $U(1)$ group.
Since we want to study additional structural symmetries of the NHDM scalar sectors, 
we disregard such common phase rotations. 
Thus, we are interested not in $\tilde G \subseteq U(N)$ but in symmetry groups 
$G = \tilde G/U(1) \subseteq U(N)/U(1) \simeq PSU(N)$.

We draw the readers attention to the fact that the traditionally used $SU(N)$,
which is obtained by imposing $\det U = 1$,
does not eliminate the above common phase rotations completely.
Indeed, the group $SU(N)$ contains a non-trivial center $Z(SU(N))$,
that is, the $\Z_N$ subgroup of the phase rotations $\exp(2\pi i k /N)\cdot \id_N$, $k = 0, 1, \dots, N-1$. 
Transformations from the center act trivially on all Higgs lagrangians
and do not offer any insight into the structural properties of the Higgs sector.
Therefore, we are led again to the factor group $SU(N)/Z(SU(N)) \simeq PSU(N)$.
A detailed discussion of these matters can be found in \cite{Ivanov:2011ae}.

It is well known that any abelian subgroup of $U(N)$ or $SU(N)$ can be represented, in a suitable basis, 
as pure phase rotations. They also correspond to phase rotations upon the homomorphism $SU(N) \to SU(N)/Z(SU(N)) \simeq PSU(N)$.
However there exists an additional abelian group in $PSU(N)$, which cannot be represented as phase
rotations because its full pre-image inside $SU(N)$ is non-abelian, 
$(\Z_N \times \Z_N)\rtimes \Z_N$. Once factored by the center $Z(SU(N))$, 
it produces the abelian group $\Z_N \times \Z_N \subset PSU(N)$, see again \cite{Ivanov:2011ae}.

When determining the symmetry group $G$ of a given NHDM potential, one must identify its full symmetry content. 
In this way, one must avoid the situations in which accidental symmetries appear which are not included in $G$.
To give a simple example, one can construct a 2HDM model with the global symmetry $\Z_p$ with any $p > 2$.
However the scalar potential of this model will be automatically invariant under the accidental continuous global symmetry $U(1)$,
which of course contains any $\Z_p$ as a subgroup.
The total symmetry content of such a model is $U(1)$, not $\Z_p$; 
labeling this model as $\Z_p$-invariant 2HDM would be a misnomer.
Thus, whenever we say that an NHDM has the symmetry group $G$, 
we always make sure that it does not possess a larger Higgs family symmetry group which would contain $G$ as a proper subgroup.
Following the notation of \cite{Ivanov:2011ae,Ivanov:2012fp}, we call such groups realizable
as there exist NHDM examples whose full Higgs family symmetry content is given by $G$.

It may also happen that the Higgs potential possesses generalized $CP$ symmetries, that is,
it remains invariant under the global transformations of the type $\phi_i(\vec r, t) \mapsto X_{ij} \phi_j^*(-\vec r, t)$,
where $X_{ij}$ is again a unitary matrix \cite{Ecker:1981wv,Ecker:1983hz,Ecker:1987qp,Grimus:1995zi,Branco:1999fs}.
In particular, if it happens that all coefficients of the Higgs potential are real in some basis,
the scalar sector obviously possesses the usual $CP$ symmetry, the one corresponding to $X = \id_N$.
However it is also possible to construct NHDM potentials which 
are invariant under a higher-order $CP$ symmetry; such a model is still explicitly $CP$ invariant
in spite of having complex free parameters in any basis.
This opportunity leads to a new class of the 3HDM based on CP4, the generalized $CP$ symmetry of order 4, 
which was clearly demonstrated in \cite{Ivanov:2015mwl,Haber:2018iwr}.
Higher order $CP$ symmetries in NHDM with $N > 4$ were identified in \cite{Ivanov:2018qni}.

\subsection{Finite symmetry groups in the 3HDM scalar sector}

We now outline the procedure which was used in 
\cite{Ivanov:2011ae,Ivanov:2012fp,Ivanov:2012ry} to classify all the finite symmetry groups of the 3HDM scalar sector.

Let us first remind the reader that, when discussing finite groups, the word ``order'' has two distinct meanings
depending on whether it is applied to an element of the group or the group itself.
The order of a finite group $G$, denoted as $|G|$, 
is the number of elements in $G$. The order of an element $g \in G$ is the smallest positive integer $n$ such that 
$g^n = e$, the identity of the group.

If a group has a proper subgroup $H \subset G$, its order must, by Lagrange’s theorem, divide
the order of the group: 
$|H|$ divides $|G|$. If proper subgroups
exist, some of them must be abelian. 
A simple way to obtain
an abelian subgroup is to pick up an element $g \in G$ and consider its successive powers $g^k$.
If the order of the element $g$ is $n$, we will get
the cyclic group $\Z_n \subset G$.

The inverse of Lagrange’s theorem is not, generally
speaking, true: namely, if $p$ is a divisor of $|G|$, the group $G$
is not required to posses a subgroup of order $p$ (for example, the group $A_4$ of order 12 does not possess subgroups of order 6). 
However,
if $p$ is a prime divisor of $|G|$,
then, according to Cauchy’s theorem, 
such a subgroup must
exist and is equal to $\Z_p$ (as there are no other groups of
prime order). 

Suppose we know the full list of finite abelian groups $A$ which can appear in a certain class of models
and we are looking for finite non-abelian group $G$ in this model. Then, based on the above basic facts,
we can conclude that the order of $|G|$ must contain only those prime factors which appear in the orders of $|A|$.
In the case of the 3HDM, 
the full list of finite abelian subgroups $A \subset PSU(3)$ which can be used for 3HDM scalar sector is
\begin{equation}
\mbox{$A$ in 3HDM:}\qquad \Z_2\,,\quad 
\Z_3\,,\quad 
\Z_4\,,\quad 
\Z_2\times \Z_2\,,\quad 
\Z_3\times \Z_3\,.
\label{3HDM-abelian}
\end{equation}
The first four groups here can be represented as phase rotation groups, while the last one 
arises as $\Delta(27)/\Z_3 \simeq \Z_3\times \Z_3$, where $\Delta(27) \subset SU(3)$ and the factored $\Z_3 = Z(SU(3))$.
We see that the orders of all these abelian groups contain only primes: 2 and 3.
Therefore, the order of any non-abelian finite group $G$ must contain only these two primes: $|G| = 2^a 3^b$.
Then according to Burnside’s $p^aq^b$-theorem, the group $G$ is solvable (Theorem 7.8 in \cite{Isaacs});
for a physicist-friendly introduction to solvable groups, see section~3 of \cite{Ivanov:2012fp}. 

A solvable group contains a normal abelian subgroup $A$ so that one can define the factor group $G/A$.
This information by itself is not sufficient to limit the order of $|G|$ and deduce the structure of the factor group $G/A$.
However it turns out that in the 3HDM one can prove a stronger statement:
any finite non-abelian group $G$ must contain a normal {\em maximal} abelian subgroup \cite{Ivanov:2012fp}.
This additional piece of information represents the key step in the procedure because
existence of a normal maximal abelian subgroup $A$ implies that $G/A \subseteq \Aut(A)$,
the automorphism group of $A$.
Thus, one arrives at a systematic procedure to classify all realizable groups $G$: 
\begin{itemize}
\item Take $A$ from the list Eq.~\eqref{3HDM-abelian}, compute its automorphism group $\Aut(A)$, 
and list all subgroups $K \subseteq \Aut(A)$.
\item
If $G/A \simeq K$, then $G$ can be constructed as an extension of $A$ by $K$.
\item
In general, there are two types of extensions. The so-called split extension, also known as semi-direct product
$A \rtimes K$, implies that $G$ contains, among its subgroups, not only $A$ but also $K$.
The non-split extension denoted as $A\,.\,K$ implies that $G$ does not contain $K$,
or, in other words, that $K$ is not closed under multiplication when embedded in $G$.
We will see explicit examples of this situation in the next section.
\item
Even if one takes specific $A$ and specific $K$, the extension is not unique. 
One needs to construct all the cases explicitly.
\item
By checking all $A$'s, all $K$'s, and performing all possible extensions, one obtains the full list 
of finite non-abelian groups $G$ for the 3HDM scalar sector.
\end{itemize}
This procedure was performed in \cite{Ivanov:2012fp,Ivanov:2012ry} and 
produced the following full list of finite non-abelian groups $G$ 
available in the 3HDM which do not lead to accidental symmetries:
\begin{equation}
\mbox{$G$ in 3HDM:}\qquad S_3\,,\quad 
D_4\,,\quad 
A_4\,,\quad 
S_4\,,\quad 
\Delta(54)/\Z_3\,,\quad 
\Sigma(36)\,.
\label{3HDM-non-abelian}
\end{equation}
Trying to build a 3HDM scalar sector on any other Higgs family symmetry group absent from the lists 
Eq.~\eqref{3HDM-abelian} and Eq.~\eqref{3HDM-non-abelian}
will unavoidably lead to a continuous accidental symmetry.


\section{Symmetry options for the 4HDM: the general strategy}\label{section-4HDM-general}

\subsection{Abelian groups in the 4HDM}

In the present work, we apply the above strategy to the 4HDM.
As we will shortly see, this procedure will not exhaust all non-abelian symmetry groups which are realizable
for the 4HDM scalar sector. However it will still generate an important class of symmetry groups
and lead to symmetry-based 4HDMs, some of which have never been studied before.

The full list of abelian symmetry groups in the 4HDM scalar sector was already established in \cite{Ivanov:2011ae}.
Focusing on finite rephasing symmetry groups, we find here all abelian groups of order at most 8:
\begin{equation}
\mbox{$A$ in 4HDM:}\qquad \Z_k\ \mbox{with}\ k = 2,\dots,8\,,\quad 
\Z_2\times \Z_2\,,\quad 
\Z_4\times \Z_2\,,\quad 
\Z_2\times \Z_2 \times \Z_2\,.
\label{4HDM-abelian-1}
\end{equation}
In addition, we have a special abelian group
\begin{equation}
\Z_4 \times \Z_4 \subset PSU(4)\,,\label{4HDM-abelian-2}
\end{equation}
whose full pre-image in $SU(4)$ is the non-abelian group $(\Z_4 \times \Z_4)\rtimes \Z_4$ generated by
\begin{equation}
a = \sqrt{i}\left(\begin{array}{cccc} 
1 & 0 & 0 & 0\\ 
0 & i & 0 & 0\\ 
0 & 0 & -1 & 0\\ 
0 & 0 & 0 & -i\\ 
\end{array}\right)\,,\quad
b = \sqrt{i}\left(\begin{array}{cccc} 
0 & 1 & 0 & 0\\ 
0 & 0 & 1 & 0\\ 
0 & 0 & 0 & 1\\ 
1 & 0 & 0 & 0\\ 
\end{array}\right)\,.\label{generators-Z4Z4}
\end{equation}
The factor $\sqrt{i}$ is introduced in $a$ and $b$ to make sure that $\det a = \det b = 1$.
One can immediately verify that the the two generators do not commute inside $SU(4)$ but
their commutator $[a,b]\equiv aba^{-1}b^{-1} = -i\cdot \id_4$ belongs to the center of $SU(4)$.
Therefore, factoring this group by the center of $SU(4)$ produces the abelian group of Eq.~\eqref{4HDM-abelian-2}.

The orders of the abelian groups from the lists Eq.~\eqref{4HDM-abelian-1} and Eq.~\eqref{4HDM-abelian-2} 
have prime factors 2, 3, 5, and 7.
Thus, the order of any non-abelian group $G$ can only have these prime factors in its prime decomposition.
However the fact that we now have four different primes available renders Burnside's $p^aq^b$ theorem
inapplicable. As a result, there is no guarantee that any $G$ possesses a normal abelian subgroup
let alone a normal maximal abelian subgroup. Thus, although we still can apply the same procedure
that worked so nicely for the 3HDM (extending abelian groups by their automorphisms),
there is no guarantee that it will exhaust all the finite non-abelian groups in the 4HDM scalar sector.

\begin{table}[H]
\centering
\begin{tabular}[t]{cc}
\toprule
$A$ & $\Aut(A)$ \\
\midrule
$\Z_2$ & $\{e\}$ \\
$\Z_3$ & $\Z_2$ \\
$\Z_4$ & $\Z_2$ \\
$\Z_5$ & $\Z_4$ \\
$\Z_6$ & $\Z_2$ \\
$\Z_7$ & $\Z_6$ \\
$\Z_8$ & $\Z_2\times \Z_2$ \\
\bottomrule
\end{tabular}
\quad
\begin{tabular}[t]{cc}
\toprule
$A$ & $\Aut(A)$ \\
\midrule
$\Z_2 \times \Z_2$ & $S_3$ \\
$\Z_2 \times \Z_4$ & $D_4$ \\
$\Z_2 \times \Z_2 \times \Z_2$ & $SL(3,2)\simeq PSL(2,7)$ \\
$\Z_4 \times \Z_4$ & SmallGroup(96,195) \\
\bottomrule
\end{tabular}
\caption{The list of all finite abelian symmetry groups $A$ of the 4HDM scalar sector
and their automorphism groups $\Aut(A)$.}
\label{table-abelian}
\end{table}

Still, we are going to pursue this strategy, and for this purpose, we need to know
the automorphism groups for all the abelian group Eq.~\eqref{4HDM-abelian-1} and Eq.~\eqref{4HDM-abelian-2}.
In Table~\ref{table-abelian}, we provide this list.
For the case of $\Z_4\times\Z_4$, we identified its automorphism group using the computer algebra system {\tt GAP}~\cite{GAP}
which provides access to the {\tt SmallGroups} library~\cite{SmallGroups}.
For the other groups, their automorphism groups are calculated directly using elementary group theoretic methods.

In this work, we will only deal with extensions of the cyclic groups (the left table).
The task of building extensions of products of cyclic groups (the right table) and finding the corresponding constraints
on the scalar potential is much more involved and brings in several challenges.
First, products of cyclic groups have more than one generator. When extending the group by its automorphisms, 
we need to specify how each automorphism acts on all the generators.
Even for a fixed group $A$ and a fixed subgroup of the automorphism group $H \subseteq \Aut(A)$, 
this action is often not unique, and the number of cases can be large for products of cyclic groups.
Second, as we will see below, there are several non-equivalent ways the cyclic groups $\Z_2$ and $\Z_4$
can be imposed on the 4HDM potential. The extensions available depend in a significant way on the particular implementation.
Third, the automorphism groups in Table~\ref{table-abelian}, right, are large and have many subgroups.
For example, $\Aut(\Z_2\times\Z_2\times\Z_2)\simeq SL(3,2)$ is a group of order 168 and contains 179 subgroups.
Since we aim at a complete classification, we need to deal, in principle, with all of these subgroups.
Fourth, the group $\Z_4 \times \Z_4$ is generated not only by phase rotations but also by a cyclic permutation,
see Eq.~\eqref{generators-Z4Z4}.
Extending it by its automorphisms will involve matrices which are not limited to permutations
but will mix all Higgs doublets. The analysis of this situation requires special treatment.

All these technical complications can be overcome, but the sheer amount of work forces us to delegated this analysis to a follow-up paper.

\subsection{Working in $PSU(4)$}\label{subsection-working-PSU4}

Consider a cyclic group $\Z_n$ with any integer $n > 1$.
It is generated by a transformation $a$ such that $a^n = e$ and 
$a^m \not = e$ for $0< m < n$. 
Its four-dimensional representation acting on the four Higgs doublets can be parametrized, in a suitable basis, by the diagonal matrices
\begin{equation}
a_{U(4)} = \diag(\eta^{q_1},\, \eta^{q_2},\, \eta^{q_3},\, \eta^{q_4})\,, \quad \eta \equiv e^{2i\pi/n}\,, \quad \eta^n = 1\,.
\label{generator-U4}
\end{equation}
Here, $q_i$ are known as $\Z_n$ charges which are integers modulo $n$.

As discussed before, the group $U(4)$ contains the global $U(1)$ of common phase shifts of all four doublets, 
which is included in the local $U(1)_Y$ gauge group. 
When constructing the Higgs potential, we do not distinguish the transformations which differ by this overall transformation.
Thus, our interest lies in finding symmetry groups which are subgroups of $U(4)/U(1)\simeq PSU(4)$, not $U(4)$.

The first step towards $PSU(4)$ is to require $\det a = 1$, which leads us to $SU(4)$.
However, $SU(4)$ contains a non-trivial center: $Z(SU(4))\simeq \Z_4$ generated by the simultaneous multiplication of all doublets by $i$.
Factoring by this center is the second step which brings us to $SU(4)/\mathbb{Z}_4\simeq PSU(4)$.  

However, working directly in $PSU(4)$ is not convenient.
If $\Phi = (\phi_1,\phi_2,\phi_3,\phi_4)$ is a vector in the space of four doublets, 
then the elements of $PSU(4)$ act not on $\Phi$ but on its $\Z_4$ orbit
$\Z_4\Phi = \{\Phi,\, i\Phi,\, -\Phi,\, -i\Phi\}$.
Another way to state this property is to say that multiplication by $i$ has no effect: 
we just reshuffle the elements inside the orbit
but the orbit itself is unchanged. 
From the computational point of view, it is much more convenient to 
define a transformation from $PSU(4)$ by choosing its representative transformation
$a \in SU(4)$ but remembering that matrices $a$, $ia$, $-a$, $-ia$ 
represent in fact the same transformation within $PSU(4)$ and are considered equivalent. 

With these remarks, we fix our conventions as follows. 
The generator $a$ of the group $\Z_n \in PSU(4)$ is written via 
its representative phase rotation transformation $a\in SU(4)$ in the form
\begin{equation}
a = \eta^{r/4} \cdot \diag(\eta^{q_1},\, \eta^{q_2},\, \eta^{q_3},\, \eta^{q_4})\,, \quad \eta \equiv e^{2i\pi/n}\,, \quad \eta^n = 1\,,
\label{generator-PSU4}
\end{equation}
which is subject to a milder constraint: $a^n = i^r\mathbf{1}_{4}$.
The $\Z_n$ charges $q_k$ are integer, and the value of $r$ is determined from the requirement that $\det a = 1$,
which translates into $r + \sum_{k=1}^4 q_k = 0$ mod $n$.

\subsection{Split and non-split extensions}\label{subsection-split}

We will build finite non-abelian groups with the aid of a procedure called the group extension.
Let us briefly explain it and illustrate it with basic examples;
more details can be found in \cite{Ivanov:2012fp}.

A subgroup $H$ in $G$ is called normal (denoted as $H \lhd G$) if $g^{-1}Hg = H$ for all $g\in G$.
A normal subgroup allows us to define the group structure on the set of cosets of $H$, 
which is now called the factor group $G/H$. 
Thus, normal subgroups help break the group
into two smaller groups, which simplifies its study.

The group-constructing procedure inverse to factoring is called {\em extension}.
Given two groups, $N$ and $H$, a group $G$ is called an extension of $N$ by $H$ 
if there exists $N_0 \lhd G$ such that $N_0 \simeq N$ and $G/N_0 \simeq H$.
In the case when, in addition, $H$ is also isomorphic to a subgroup of $G$ and $G=NH$,
we deal with the split extension. The criterion for $G$ to be a split extension
can also be written as existence of $N \lhd G$ and $H \leq G$ such that $NH = G$ and $N\cap H = 1$,
so that $G/N = H$.
The group $G$ is then called a semidirect product $G = N \rtimes H$.
An extension which is not split is called a non-split extension and is denoted as $G = N\,.\,H$. 
Notice that even if two groups $N$ and $H$ are fixed, they can support several extensions and split extensions.

For the most elementary example, consider extensions of $N=\Z_2$ (generated by $a$) by $H=\Z_2$ (generated by $b$),
which should produce a group of order 4.
For a split extension, we need a group $G$ which contains two distinct subgroups
isomorphic to $N$ and $H$. The only choice is $G = \Z_2\times \Z_2$, which can be presented as $\langle a,b\, |\,
a^2=b^2=(ab)^2=e\rangle$.
For a non-split extension, we require that only $N$ is isomorphic to a subgroup of $G$.
Thus, we still have $a^2 = e$, while $b^2$ must be different from the unit element,
that is, $H$ is not closed inside $G$ under multiplication. 
In this example, we have to set $b^2 = a$ producing the group $\Z_4$.
So, $\Z_4$ does not split over $\Z_2$, while $\Z_2\times \Z_2$ does.
Both of them, however, can be factored by $N$ and give $G/N \simeq H = \Z_2$.

For a less elementary but very useful example, consider non-abelian extensions of 
$N = \Z_4$ generated by $a$ by $H = \Z_2$ generated by $b$.
Since $N$ is normal in the group $G$, we have $b^{-1}Nb = N$.
Since we aim at non-abelian $G$, the only choice is $b^{-1}ab = a^{-1}$.
However we also need to specify $b^2$ inside the group $G$. We have two options:
\begin{itemize}
\item 
$b^2 = e$. Then, $G$ contains a copy of $H$, which does not intersect with $N$, and we obtain a split extension.
This is the semidirect product $\Z_4 \rtimes \Z_2 \simeq D_4$, the symmetry group of the square.
\item 
$b^2 \not = e$. The only choice is then $b^2 = a^2$ (assigning $b^2 = a$ would produce the abelian group $\Z_8$). 
In this way we obtain a non-split extension,
and the group is known as $\Z_4\,.\,\Z_2 \simeq Q_4$, the quaternion group.
\end{itemize}
Both groups are of order 8, they possess the normal subgroup $\Z_4$, and their factor groups by $\Z_4$ gives $\Z_2$.
Still $D_4$ and $Q_4$ are distinct; for example, $D_4$ contains a subgroup $\Z_2\times \Z_2$, while $Q_4$ does not. 

It is interesting to note that, while trying to extend $\Z_4$ by $\Z_2$ within the 3HDM, 
one can arrive at the $D_4$-invariant 3HDM but cannot produce a realizable $Q_4$-invariant 3HDM,
see details in \cite{Ivanov:2012fp}.
The difference is that the generator $b$ of the form of $b^2 = a^2$ leads to more constraints on the scalar potential
than $b^2=e$, and, as a result, the desired $Q_4$-symmetric model acquires an accidental continuous symmetry.
As we will see below, within the 4HDM, the $Q_4$-invariant models are possible without accidental symmetries.

\subsection{The scalar potential}\label{subsection-V0}

Since the abelian groups $A$ which we discuss in this work are represented, in a suitable basis, by pure phase rotations,
let us break the $A$-symmetric scalar potential of the 4HDM into two part: the rephasing-insensitive part $V_0$, 
invariant under all phase rotations of individual doublets, 
and the rephasing-sensitive part $V(A)$ which stays invariant only
under the phase rotations which form the group $A$.
The rephasing-insensitive part $V_0$ can be written in the following form:
\begin{eqnarray}
V_0 &=& \sum_{i=1}^4 \left[m_{ii}^2\fdf{i}{i} + \Lambda_{ii}\fdf{i}{i}^2 \right]+ 
\sum_{i < j} \left[\Lambda_{ij}\fdf{i}{i}\fdf{j}{j} + \tilde \Lambda_{ij}\fdf{i}{j}\fdf{j}{i}\right]\,.\label{V0-general}
\end{eqnarray}
In total, this potential has 20 real free parameters.
The rephasing-sensitive part $V(A)$ depends on the group $A$ and will be given for every choice of $A$.

When we extend the group $A$ by its automorphisms, we will search for such transformations $b$ which
satisfy $b^{-1}A b = A$. Such transformations can be represented by certain permutations of the four doublets,
accompanied perhaps by additional phase factors.
The requirement of invariance under $b$ applies to $V_0$ as well as to $V(A)$.
Depending on which doublets are permuted under $b$, we will have one of the following constraints on $V_0$. 
\begin{itemize}
\item 
The permutation $\phi_1 \leftrightarrow \phi_2$ leads to the following constraints:
\begin{equation}
m_{11}^2 = m_{22}^2\,, \quad \Lambda_{11} = \Lambda_{22}\,, \quad \Lambda_{1k} = \Lambda_{2k}\,, 
\quad \tilde\Lambda_{1k} = \tilde\Lambda_{2k}\,, \quad k = 3,4.
\label{V0-2-1-1}
\end{equation} 
The potential $V_0$ then has 14 free parameters left.
\item 
The permutation $\phi_1 \leftrightarrow \phi_2$ and, simultaneously, $\phi_3 \leftrightarrow \phi_4$, leads to the following constraints:
\begin{eqnarray}
&& m_{11}^2 = m_{22}^2\,, \quad m_{33}^2 = m_{44}^2\,, \quad \Lambda_{11} = \Lambda_{22}\,, \quad \Lambda_{33} = \Lambda_{44}\,, \nonumber\\
&&\Lambda_{13} = \Lambda_{24}\,, \quad \Lambda_{14} = \Lambda_{23}\,, \quad 
\tilde\Lambda_{13} = \tilde\Lambda_{24}\,, \quad \tilde\Lambda_{14} = \tilde\Lambda_{23}\,.
\label{V0-2-2}
\end{eqnarray} 
The potential $V_0$ has only 12 free parameters left in this case.
\item 
If we build a 4HDM with three Higgs doublets transforming as a triplet under some discrete group,
we encounter the cyclic permutation of three doublets $\phi_1 \mapsto \phi_2\mapsto \phi_3 \mapsto \phi_1$. 
It leads to the following constraints:
\begin{eqnarray}
&& m_{11}^2 = m_{22}^2 = m_{33}^2\,, \quad \Lambda_{11} = \Lambda_{22} = \Lambda_{33}\,, 
\quad \Lambda_{12} = \Lambda_{23} = \Lambda_{13}\,, 
\quad \tilde\Lambda_{12} = \tilde\Lambda_{23}  = \tilde\Lambda_{13}\,, \nonumber\\
&&
\Lambda_{14} = \Lambda_{24} = \Lambda_{34}\,, 
\quad \tilde\Lambda_{14} = \tilde\Lambda_{24}  = \tilde\Lambda_{34}\,.
\label{V0-3-1}
\end{eqnarray} 
In this case, the number of the free parameters of $V_0$ is reduced to 8.
Notice that the potential $V_0$ then automatically becomes invariant under all permutations of the first three doublets
(the group $S_3$), not only the cyclic ones.
\item 
Finally, if we encounter the cyclic permutation of all four doublets 
$\phi_1 \mapsto \phi_4\mapsto \phi_2 \mapsto \phi_3 \mapsto \phi_1$, we arrive at the following constraints:
\begin{eqnarray}
&& m_{11}^2 = m_{22}^2 = m_{33}^2 = m_{44}^2 \equiv m^2\,, \quad 
\Lambda_{11} = \Lambda_{22} = \Lambda_{33} = \Lambda_{44} \equiv \Lambda\,,
\nonumber\\
&& 
\Lambda_{13} = \Lambda_{14} = \Lambda_{23} = \Lambda_{24} \equiv \Lambda'\,, 
\quad 
\Lambda_{12} = \Lambda_{34} \equiv \Lambda''\,,\nonumber\\
&&
\tilde\Lambda_{13} = \tilde\Lambda_{14}  = \tilde\Lambda_{23}  = \tilde\Lambda_{24} \equiv \tilde\Lambda'\,,
\quad 
\tilde \Lambda_{12} = \tilde \Lambda_{34} \equiv \tilde\Lambda''\,.
\label{V0-4-0}
\end{eqnarray} 
In this case, the potential $V_0$ contains only 6 real free parameters:
\begin{eqnarray}
V_0 &=& m^2 \left(\fdfn{1}{1} + \fdfn{2}{2} + \fdfn{3}{3} + \fdfn{4}{4}\right)
+ \Lambda \left(\fdfn{1}{1} + \fdfn{2}{2} + \fdfn{3}{3} + \fdfn{4}{4}\right)^2\label{V0-4D}\\
&&+\Lambda' \left(\fdfn{1}{1} + \fdfn{2}{2}\right)\left(\fdfn{3}{3} + \fdfn{4}{4}\right) 
+\Lambda'' \left[\fdf{1}{1}\fdf{2}{2} + \fdf{3}{3}\fdf{4}{4}\right]\nonumber\\
&& + \tilde\Lambda' \left[|\fdf{1}{3}|^2 + |\fdf{2}{3}|^2 + |\fdf{1}{4}|^2 + |\fdf{2}{4}|^2 \right]
+ \tilde\Lambda''\left[|\fdf{1}{2}|^2 + |\fdf{3}{4}|^2\right]\,.\nonumber
\end{eqnarray}
In addition to cyclic permutations, it is also automatically invariant 
under the simultaneous exchange $\phi_1 \leftrightarrow \phi_4$ and $\phi_2 \leftrightarrow \phi_3$.
All these transformations form the group $D_4$, the symmetry group of the square
whose vertices are labeled by the four doublets.
\end{itemize}
This list of options is complete up to renaming of doublets.


\section{Non-abelian extensions of cyclic groups}\label{section-4HDM-cases}

\subsection{Extending $\Z_8$}

\subsubsection{The $\Z_8$-invariant 4HDM}

We begin our analysis with the abelian group $\Z_8$ and its possible non-abelian extensions.
In this first example, we will expose the procedure step by step, 
and in the subsequent examples, we will rely on this sequence of steps.

The starting point is to write the 4HDM invariant under the symmetry group $\Z_8$. 
Following the discussion in section~\ref{subsection-working-PSU4},
we represent the generator of this group by an $SU(4)$ phase rotation matrix $a$ which, in a suitable basis, 
has the following form:
\begin{equation}
a = \eta^{1/4}\, \cdot \diag(\eta,\, \eta^2,\, \eta^4,\, 1)\,, \quad \eta \equiv e^{i\pi/4} = \sqrt{i}\,, \quad \eta^8 = 1\,.
\label{Z8-a3}
\end{equation}
We draw the reader's attention to the fact that the choice of the $\Z_8$ charges used here is not arbitrary and is,
in fact, unique up to the doublets permutation and the possible $i = \eta^2$ factors.
These charges are fixed by the Smith normal form technique developed in \cite{Ivanov:2011ae} and briefly summarized in Appendix~\ref{appendix-SNF}. 
To double check that we do not miss other charge assignments, a Python code 
{\tt 4HDM Toolbox} was written, which checks all possible 
monomial combinations and identifies the rephasing symmetry group and its charge assignments, 
see details in Appendix~\ref{appendix-SNF}. This code confirmed the uniqueness of the $\Z_8$ charge choice.

The Higgs potential invariant under so-defined $\Z_8$ contains, apart from the rephasing invariant terms $V_0$ given in Eq.~\eqref{V0-general}, 
the following rephasing-sensitive terms: 
\begin{equation}
V({\Z_8}) = \lambda_{1} \fdf{2}{1} \fdf{4}{1} + \lambda_{2} \fdf{3}{2} \fdf{4}{2} + \lambda_{3} \fdf{4}{3}^2 + h.c.
\label{VZ8}
\end{equation}
Here, all the coefficients can, in principle, be complex. However upon a suitable rephasing of the four doublets, 
all $\lambda_i$ can be made real, without affecting the symmetry group $\Z_8$. 
We conclude that imposing $\Z_8$ on the scalar sector of the 4HDM automatically leads to explicit $CP$ conservation.
For generic values of the coefficients, this potential has no other symmetries.

\subsubsection{The automorphism group of $\Z_8$}

If $a$ generates $\Z_8$, then $a^3$, $a^5$, and $a^7 = a^{-1}$ can also play the role of its generator.
Therefore, the transformation $b$ which maps $a\mapsto a^3$ and the transformation $c$ which maps $a\mapsto a^{-1}$
upon conjugation
\begin{equation}
b^{-1}ab = a^3\,, \quad c^{-1}ac = a^{-1}
\end{equation}
leave the $\Z_8$ group unchanged and represent its automorphisms. 
One immediately checks that applying $b$ twice maps $a$ to $a^9 = a$. 
Therefore, $b^2$ is the trivial automorphism, and so is $c$.
Finally, applying $bc$ (or $cb$) maps $a\mapsto a^5$. Clearly, $(bc)^2$ is also the trivial automorphism.
Thus, we find that $\Aut(\Z_8) \simeq \Z_2\times \Z_2$ generated by $b$ and $c$.

\subsubsection{Attempts at extending $\Z_8$}

Let us consider the split extension (semidirect product) 
\begin{equation}
\Z_8 \rtimes \Z_2 = \langle a, b \,|\, a^8 = e, b^2 = e, b^{-1} a b = a^3\rangle\,.\label{extension-Z8Z2-1}
\end{equation}
We want to represent $b$ as a matrix in $SU(4)$ taking into account that $i^r$ factors can always accompany
the identity in matrix equations. 
Therefore, we need $b$ to solve the matrix equation
\begin{equation}
ab = ba^3\cdot i^r
\end{equation}
with any integer $r$. Noting that $i = \eta^2$ and writing this matrix equation explicitly, we get
\begin{equation}
\eta^{1/4}
\left(\!\! \begin{array}{cccc} 
\eta\,{\gray b_{11}} & \eta\,{\gray b_{12}} & \eta\,{\gray b_{13}} & \eta\,{\gray b_{14}}\\ 
\eta^2{\gray b_{21}} & \eta^2{\gray b_{22}} & \eta^2{\gray b_{23}} & \eta^2{\gray b_{24}}\\ 
\eta^4{\gray b_{31}} & \eta^4{\gray b_{32}} & \eta^4{\gray b_{33}} & \eta^4{\gray b_{34}}\\ 
{\gray b_{41}} & {\gray b_{42}} & {\gray b_{43}} & {\gray b_{44}}\\
\end{array}\!\!\right)
= \eta^{3/4} \eta^{2r}
\left(\!\! \begin{array}{cccc} 
\eta^3{\gray b_{11}} & \eta^6{\gray b_{12}} & \eta^4{\gray b_{13}} & {\gray b_{14}}\\ 
\eta^3{\gray b_{21}} & \eta^6{\gray b_{22}} & \eta^4{\gray b_{23}} & {\gray b_{24}}\\ 
\eta^3{\gray b_{31}} & \eta^6{\gray b_{32}} & \eta^4{\gray b_{33}} & {\gray b_{34}}\\ 
\eta^3{\gray b_{41}} & \eta^6{\gray b_{42}} & \eta^4{\gray b_{43}} & {\gray b_{44}}\\
\end{array}\!\!\right)\,.
\label{extension-Z8Z2-2}
\end{equation}
We need a non-trivial solution to this matrix equation on $b$, at least for some $r$.
Since the entries of $b$ are the same on both sides of the equality,
we compare the two matrices element by element and
either equate the rephasing factors or set the corresponding entry of $b$ to zero.
For example, the factors in front of $b_{11}$ are $\eta^{5/4}$ in the left hand side and $\eta^{3+2r+3/4}$ in the right hand side.
Since they cannot be made equal by any integer $r$, we must set $b_{11} = 0$.

A quick inspection reveals that all the elements must be set to zero because of the mismatch 
between the coefficients $\eta^{1/4}$ and $\eta^{3/4}$ which cannot be compensated by any integer power of $\eta$.
Therefore, the extension given by Eq.~\eqref{extension-Z8Z2-1}
cannot be realized within the 4HDM.

The second option is to extend $\Z_8$ to $\Z_8\rtimes \Z_2$ using $c$. 
This leads to the matrix equation $ac = ca^{-1}\cdot i^r$. Writing it in the same way,
we again notice the mismatch between $\eta^{1/4}$ and $\eta^{-1/4}$ which cannot be compensated by
any integer power of $i$. Thus, this extension does not fit the 4HDM.

Finally, we try to extend $\Z_8$ to $\Z_8\rtimes \Z_2$ using $d \equiv bc$:
\begin{equation}
\Z_8 \rtimes \Z_2 = \langle a, d \,|\, a^8 = e, d^2 = e, d^{-1} a d = a^5\rangle\,.\label{extension-Z8Z2-3}
\end{equation}
Now, the matrix equation $ad = da^5\cdot i^r$ does not run into the above problem because $\eta^{1/4}$
and $\eta^{5/4}$ do differ by a single power of $\eta$.
However, there is an insufficient number of non-zero elements of $b_{ij}$ for any value of $r$, 
making the matrix $d$ non-invertible. For example, for $r=0$, only $b_{14} \not = 0$, while all other elements must be set to zero.
Therefore, this choice does not represent a solution of $d^{-1} a d = a^5$.

We conclude that none of the possible (split) extensions of $\Z_8$ by its automorphisms 
can lead to a viable symmetry group in the 4HDM scalar sector.
Dropping the assumption $b^2=e$ 
and passing to non-split extensions, $b^2 \in \Z_8$, does not help:
the problems exposed above do not depend on the form of $b^2$.
The origin of the obstacle is the very particular way the $\Z_8$ group is embedded in $PSU(4)$.

\subsection{Extending $\Z_7$}

\subsubsection{The $\Z_7$-invariant 4HDM}

As before, we begin by constructing the 4HDM potential based on the symmetry group $\Z_7$.
According to \cite{Ivanov:2011ae}, there is a basis in which its generator $a$ acts 
on the four Higgs doublets as
\begin{equation}
a = \diag(\eta,\, \eta^2,\, \eta^4,\, 1)\,, \quad \eta \equiv e^{2\pi i/7}\,, \quad \eta^7 = 1.
\label{Z7-a1}
\end{equation}
This choice is unique up to multiplication by $i$, doublet permutations, and automorphisms of the $\Z_7$ group.
The Higgs potential invariant under this symmetry is defined, in addition to $V_0$ of Eq.~\eqref{V0-general}, 
by the following rephasing-sensitive terms:
\begin{equation}
V({\Z_7}) = \lambda_{1} \fdf{2}{1} \fdf{4}{1} + \lambda_{2} \fdf{3}{2} \fdf{4}{2} + \lambda_{3} \fdf{1}{3} \fdf{4}{3} + h.c.
\end{equation}
It differs from $V({\Z_8})$ in Eq.~\eqref{VZ8} only by the last term.
As before, using the rephasing basis change freedom, one can make all three $\lambda_i$ real and positive,
which implies that $\Z_7$-invariant 4HDM is explicitly $CP$ conserving.

\subsubsection{The automorphism group of $\Z_7$}

The group $\Z_7$ is of prime order, and any of its non-unit elements can play the role of the generator.
Therefore, all the maps $a \mapsto a^q$, with $q = 1, \dots, 6$, are automorphisms of $\Z_7$
and form the group $\Aut(\Z_7) \simeq \Z_6$.
Since we are interested in extending $\Z_7$ not only by the full $\Aut(\Z_7)$ but also by its subgroups,
let us study separately the order-2 and the order-3 elements from $\Aut(\Z_7)$ defined by
\begin{equation}
b^{-1} a b = a^6 = a^{-1}\,, \quad 
c^{-1} a c = a^2\,. \label{Z7-bc}
\end{equation}
These two transformations commute, and their product, which maps $a\mapsto a^5$, is of order 6.

\subsubsection{Extension of $\Z_7$ by $\Z_2$}

Let us first construct the split extension
\begin{equation}
\Z_7 \rtimes \Z_2 = \langle a, b \,|\, a^7 = e, b^2 = e, b^{-1} a b = a^{-1}\rangle\,.\label{extension-Z7Z2-1}
\end{equation}
The relation between $a$ and $b$ leads to the matrix equation on $b$:
\begin{equation}
ab = ba^{-1}\cdot i^r
\end{equation}
with any integer $r$. This time, $i = \eta^{7/4}$, 
and the only chance to find non-trivial solutions to this equation is to set $r=0$.
Then, writing the matrix $b$ explicitly as in Eq.~\eqref{extension-Z8Z2-2}, we get
\begin{equation}
\left(\begin{array}{cccc} 
\eta\,{\gray b_{11}} & \eta\,{\gray b_{12}} & \eta\,{\gray b_{13}} & \eta\,{\gray b_{14}}\\ 
\eta^2{\gray b_{21}} & \eta^2{\gray b_{22}} & \eta^2{\gray b_{23}} & \eta^2{\gray b_{24}}\\ 
\eta^4{\gray b_{31}} & \eta^4{\gray b_{32}} & \eta^4{\gray b_{33}} & \eta^4{\gray b_{34}}\\ 
{\gray b_{41}} & {\gray b_{42}} & {\gray b_{43}} & {\gray b_{44}}\\
\end{array}\right)
= 
\left(\begin{array}{cccc} 
\eta^{-1}{\gray b_{11}} & \eta^{-2}{\gray b_{12}} & \eta^{-4}{\gray b_{13}} & {\gray b_{14}}\\ 
\eta^{-1}{\gray b_{21}} & \eta^{-2}{\gray b_{22}} & \eta^{-4}{\gray b_{23}} & {\gray b_{24}}\\ 
\eta^{-1}{\gray b_{31}} & \eta^{-2}{\gray b_{32}} & \eta^{-4}{\gray b_{33}} & {\gray b_{34}}\\ 
\eta^{-1}{\gray b_{41}} & \eta^{-2}{\gray b_{42}} & \eta^{-4}{\gray b_{43}} & {\gray b_{44}}\\
\end{array}\right)\,.
\end{equation}
For future convenience, let us recast this matrix equation in the form which brings in the spotlight the powers of $\eta$
in each entry: 
\begin{equation}
\left( \begin{array}{cccc} 
1 & 1 & 1 & 1\\ 
2 & 2 & 2 & 2\\ 
4 & 4 & 4 & 4\\ 
0 & 0 & 0 & 0\\ 
\end{array}\right)
\simeq 
\left( \begin{array}{cccc} 
-1 & -2 & -4 & 0\\ 
-1 & -2 & -4 & 0\\ 
-1 & -2 & -4 & 0\\ 
-1 & -2 & -4 & 0\\ 
\end{array}\right)\ \mod 7\,.
\label{extension-Z7Z2-3}
\end{equation}
We see that the only non-zero element is $b_{44}$, which means that there is no viable extension $\Z_7 \rtimes \Z_2$
in the 4HDM scalar sector.
As a consequence, the extension $\Z_7\rtimes\Z_6$ is also impossible within the 4HDM.

\subsubsection{Extension of $\Z_7$ by $\Z_3$}

Next, we try extending by $c$:
\begin{equation}
\Z_7 \rtimes \Z_3 = \langle a, c \,|\, a^7 = e, c^3 = e, c^{-1} a c = a^2\rangle\,.\label{extension-Z7Z3-1}
\end{equation}
This leads to $ac = ca^2$, where we already took into account that no $i^r$ can help find solutions. 
Tracking the powers of $\eta$ in this matrix equation, we obtain
\begin{equation}
\left( \begin{array}{cccc} 
1 & 1 & \underline{1} & 1\\ 
\underline{2} & 2 & 2 & 2\\ 
4 & \underline{4} & 4 & 4\\ 
0 & 0 & 0 & \underline{0}\\ 
\end{array}\right)
\simeq 
\left( \begin{array}{cccc} 
2 & 4 & \underline{1} & 0\\ 
\underline{2} & 4 & 1 & 0\\ 
2 & \underline{4} & 1 & 0\\ 
2 & 4 & 1 & \underline{0}\\ 
\end{array}\right)\ \mod 7\,.
\label{extension-Z7Z3-2}
\end{equation}
Now we find four elements of $c$ with matching powers of $\eta$, which are underlined in the above equation.
Therefore, the matrix $c$ has the following form:
\begin{equation}
c = \left( \begin{array}{cccc} 
0 & 0 & c_{13} & 0\\ 
c_{21} & 0 & 0 & 0\\ 
0 & c_{32} & 0 & 0\\ 
0 & 0 & 0 & c_{44}\\ 
\end{array}\right)\,. \label{extension-Z7Z3-3}
\end{equation}
In plain words, this automorphism is given by the cyclic permutation of the first three doublets
up to arbitrary phase rotations.
The phases of the non-zero elements cannot be constrained by the group-theoretic relations.
However, using the fact that we have already fixed $\lambda_{1,2,3}$ in $V(\Z_7)$ to be real and positive, 
we obtain the unique matrix $c$ by setting all its non-zero elements to be $1$.
Thus, we arrive at the 4HDM model with the symmetry group
\begin{equation}
\Z_7\rtimes \Z_3 \simeq T_7\,, \quad \mbox{generated by\ }
a = \left( \begin{array}{cccc} 
\eta & 0 & 0 & 0\\ 
0 & \eta^2 & 0 & 0\\ 
0 & 0 & \eta^4 & 0\\ 
0 & 0 & 0 & 1\\ 
\end{array}\right)\,,
\quad
c = \left( \begin{array}{cccc} 
0 & 0 & 1 & 0\\ 
1 & 0 & 0 & 0\\ 
0 & 1 & 0 & 0\\ 
0 & 0 & 0 & 1\\ 
\end{array}\right)\,,
\label{extension-Z7Z3-4}
\end{equation}
with the rephasing-sensitive potential containing only one independent coefficient:
\begin{equation}
V(T_7) = \lambda \left[\fdf{2}{1} \fdf{4}{1} + \fdf{3}{2} \fdf{4}{2} + \fdf{1}{3} \fdf{4}{3} + h.c.\right]\,.
\label{VT7-1}
\end{equation}
The rephasing-insensitive part of the potential $V_0$ must also be invariant under these permutations,
see Eq.~\eqref{V0-3-1}. 
The full Higgs potential contains only 9 real free parameters and may lead to intriguing phenomenology,
which we delegate to a future paper.

We also remark that the symmetry group $T_7$ has already received some attention in the context of neutrino masses and mixing 
model building, see e.g. \cite{Ishimori:2010au,Luhn:2007sy,Hagedorn:2008bc,Cao:2010mp,Vien:2014gza,Bonilla:2014xla}.
Here, we demonstrate that the group can also arise in the 4HDM scalar sector alone.

\subsubsection{Searching for non-split extensions of $\Z_7$}

Let us see if we can use $c$ to build a non-abelian non-split extension of $\Z_7$.
A non-split extension $G = \Z_7\,.\,\Z_3$ means that $\Z_7$ is normal in $G$
and that $G/\Z_7 \simeq \Z_3$. However we do not require $G$ to contain a copy of $\Z_3$.
This means that $c^3$ is no longer required to be $e$ but can in fact be equal to 
a non-unit element of $\Z_7$. 
The relation $c^{-1} a c = a^2$ remains intact, so that the above solution 
Eq.~\eqref{extension-Z7Z3-3} is valid in this case, too.
Now, when we calculate $c^3$, we obtain the diagonal matrix 
$c^3 = \diag(x,x,x,y)$, where $x = c_{13}c_{21}c_{32}$ and $y = c_{44}^3$.
The equality of the first three entries makes it clear that $c^3$
cannot be set equal to any of the non-trivial powers of $a$.

An alternative, and shorter, way is to notice that if $c^3$ is equal to any non-identity element of 
$\Z_7$, which is a valid generator of $\Z_7$, then all elements of $G = \Z_7\,.\,\Z_3$ 
can be presented as some power of $c$. This gives us the abelian group $\Z_{21}$.
But this group is absent in the list of viable $A$'s in the 4HDM.
In either way, we conclude that no non-split extension $\Z_7\,.\,\Z_3$ can be constructed for the 4HDM.

\subsection{Extending $\Z_6$}

\subsubsection{The two options for the $\Z_6$ generators}

The same technique of \cite{Ivanov:2011ae} allows us to find the generator $a$ of the group $\Z_6$.
However, unlike the previous two cases, the group $\Z_6$ now admits several realizations inside the 4HDM scalar sector.
Using the code {\tt 4HDM Toolbox} described in Appendix~\ref{appendix-SNF}, we found two inequivalent choices of $\Z_6$ charges:
\begin{eqnarray}
\mbox{$\Z_6$ option 1:} && a_1 = \diag(\eta,\, \eta^2,\, \eta^3,\, 1)\,, \quad \eta \equiv e^{\pi i/3}\,, \quad \eta^6 = 1 \label{Z6-a1}\\
\mbox{$\Z_6$ option 2:} && a_2 = \eta^{-1/4}\cdot\diag(\eta, \eta^2, \eta^4, 1)\label{Z6-a2}
\end{eqnarray}
That these two generators lead to different $\Z_6$ invariant models can be understood by comparing these generators cubed.
Indeed, $a_1^3 = \diag(-1, 1, -1, 1)$, which has two 2-dimensional invariant subspaces, 
while $a_2^3 = \sqrt{-i}\cdot\diag(-1, 1, 1, 1)$, which has a 3-dimensional invariant subspace. 
We also checked that other choices of the $\Z_6$ generator, which could not be brought to these ones by basis changes,
unavoidably lead to continuous accidental symmetries and are, therefore, disregarded in this study.

In the following subsections, we will extend $\Z_6$ generated by either of the two options. 
Because $\Aut(\Z_6)\simeq\Z_2$, we will construct the split extensions $\Z_6\rtimes\Z_2\simeq D_6$ and 
attempt at building the non-split extension $\Z_6\,.\,\Z_2\simeq Q_6$.

\subsubsection{Extending $\Z_6$: the first option}

In addition to the rephasing-insensitive terms $V_0$ given in Eq.~\eqref{V0-general},
the Higgs potential invariant under $a_1$ in Eq.~\eqref{Z6-a1} contains not three but five different terms:
\begin{eqnarray}
V({\Z_6}) &=& \lambda_{1} \fdf{2}{1} \fdf{4}{1} + \lambda_{2} \fdf{1}{2} \fdf{3}{2} + \lambda_{3} \fdf{4}{3}^2\nonumber\\
&& +  \lambda_{4} \fdf{1}{3} \fdf{2}{4}  + \lambda_{5} \fdf{1}{4} \fdf{2}{3} 
+ h.c.\label{V-Z6-1}
\end{eqnarray}
All the coefficients can be complex. In contrast to the previous cases, 
it is in general impossible to set all of them real by rephasing.
This implies that the $\Z_6$-invariant 4HDM can come either in the explicitly $CP$ conserving 
or $CP$ violating versions.

Notice that in this case we encounter for the first time the monomials which involve all four doublets:
$\fdf{1}{3} \fdf{2}{4}$ and $\fdf{1}{4} \fdf{2}{3}$. These two terms transform in the same way under phase rotation basis changes.
Therefore, if the coefficients in front of them, $\lambda_{4}$ and $\lambda_{5}$,
have a relative phase, it can never be eliminated by a phase rotation. 
In short, the presence of such terms, by itself, already allows the model to be $CP$ violating.

The automorphism group $\Aut(\Z_6) \simeq \Z_2$ is generated by $b$ which maps $a \mapsto a^{-1}$,
where $a$ is understood as $a_1$ in Eq.~\eqref{Z6-a1}.
This relation leads to the matrix equation 
\begin{equation}
ab = ba^{-1}\cdot i^r\label{Z6-b}
\end{equation}
with integer $0\le r < 4$. Noting that $i = \eta^{3/2}$, we translate it into
comparison of powers of $\eta$:
\begin{equation}
\left( \begin{array}{cccc} 
1 & 1 & 1 & 1\\ 
2 & 2 & 2 & 2\\ 
3 & 3 & 3 & 3\\ 
0 & 0 & 0 & 0\\ 
\end{array}\right)
\simeq 
\left( \begin{array}{cccc} 
-1 & -2 & -3 & 0\\ 
-1 & -2 & -3 & 0\\ 
-1 & -2 & -3 & 0\\ 
-1 & -2 & -3 & 0\\ 
\end{array}\right)\ + \frac{3}{2}\,r \ \mod 6\,.
\label{extension-Z6Z2-1}
\end{equation}
A non-trivial solution exists only for $r = 2$ and leads to the following matrix $b$:
\begin{equation}
b = \left( \begin{array}{cccc} 
0 & b_{12} & 0 & 0\\ 
b_{21} & 0 & 0 & 0\\ 
0 & 0 & 0 & b_{34}\\ 
0 & 0 & b_{43} & 0\\ 
\end{array}\right)\,. \label{extension-Z6Z2-2}
\end{equation}
Since $b^2 = \diag(x, x, y, y)$, with $x = b_{12}b_{21}$ and $y = b_{34}b_{43}$,
it cannot be represented by any non-zero power of $a$. 
Thus, non-split extensions are excluded.

Since $b$ is unitary, it induces the simultaneous exchanges $\phi_1\leftrightarrow\phi_2$ and $\phi_3\leftrightarrow\phi_4$,
up to possible phase shifts.
Contrary to the previous cases, we have not fixed the phases of $\lambda_i$; 
therefore, we still have the rephasing basis change freedom
to set $b_{12} = b_{21} \equiv \exp(i\alpha_{12})$ and $b_{34} = b_{43} \equiv \exp(i\alpha_{34})$,
see Appendix~\ref{appendix-trick} where the details are spelled out.
However the requirements of $\det b = 1$ and $b^2 = i^q \mathbf{1}_4$ allow us to set $\alpha_{12}=\alpha_{34}=0$ without loss of generality.
Thus, we arrive at the following viable extension of $\Z_6$:
\begin{equation}
\Z_6\rtimes \Z_2 \simeq D_6\,, \quad \mbox{generated by\ }
a = \left( \begin{array}{cccc} 
\eta & 0 & 0 & 0\\ 
0 & \eta^2 & 0 & 0\\ 
0 & 0 & \eta^3 & 0\\ 
0 & 0 & 0 & 1\\ 
\end{array}\right)\,,
\quad
b = \left( \begin{array}{cccc} 
0 & 1 & 0 & 0\\ 
1 & 0 & 0 & 0\\ 
0 & 0 & 0 & 1\\ 
0 & 0 & 1 & 0\\ 
\end{array}\right)\,.
\label{extension-Z6Z2-3}
\end{equation}
We stress once again that, although we write these transformations as $SU(4)$ matrices,
they are meant to represent the corresponding $\Z_4$ cosets.
As a result, the direct calculation shows that $b^{-1}a b = \eta^3 a^{-1} = - a^{-1}$
which belongs to the same coset as $a^{-1}$.
Therefore, $b$ indeed maps the coset $a\Z_4$ to $a^{-1}\Z_4$.

In order for the rephasing-sensitive potential $V(\Z_6)$ to be invariant under $\Z_6 \rtimes \Z_2\simeq  D_6$, 
we must require, in this basis, that 
\begin{equation}
\lambda_1 = \lambda_2\,, \quad \lambda_3 = \lambda_3^*\,,
\label{VZ6Z2-1}
\end{equation}
while $\lambda_4$ and $\lambda_5$ are unconstrained because these terms are, individually, 
invariant under $b$.
In addition, we require the rephasing-insensitive part of the potential $V_0$ 
to satisfy the conditions Eq.~\eqref{V0-2-2}.

\subsubsection{Extending $\Z_6$: the second option}

In addition to $V_0$ given in Eq.~\eqref{V0-general},
the potential invariant under $\Z_6$ generated by $a_2$ in Eq.~\eqref{Z6-a2} also contains five terms:
\begin{equation}
V_2(\Z_6) = \lambda_1\fdf{1}{3}^2 + \lambda_2\fdf{1}{2}\fdf{1}{4} + \lambda_3\fdf{2}{3}\fdf{2}{4} + 
\lambda_4\fdf{3}{2}\fdf{3}{4} + \lambda_5\fdf{4}{2}\fdf{4}{3} + h.c.
\label{V-Z6-2}
\end{equation}
It differs from Eq.~\eqref{V-Z6-1} in that it does not contain terms involving all four doublets.
We look for solutions for the matrix equation~Eq.~\eqref{Z6-b} with $a_2$. The matrix equation
for the powers of $\eta$ is:
\begin{equation}
-\frac{1}{4} + 
\left( \begin{array}{cccc} 
1 & 1 & 1 & 1\\ 
2 & 2 & 2 & 2\\ 
4 & 4 & 4 & 4\\ 
0 & 0 & 0 & 0\\ 
\end{array}\right)
\simeq 
\left( \begin{array}{cccc} 
-1 & -2 & -4 & 0\\ 
-1 & -2 & -4 & 0\\ 
-1 & -2 & -4 & 0\\ 
-1 & -2 & -4 & 0\\ 
\end{array}\right)\ + \frac{3}{2}\,r + \frac{1}{4} \ \mod 6\,.
\label{extension-Z6Z2-4}
\end{equation}
The non-trivial solution for $b$ exists for $r=1$ and leads to the following matrix:
\begin{equation}
b = \left( \begin{array}{cccc} 
b_{11} & 0 & 0 & 0\\ 
0 & 0 & 0 & b_{24}\\ 
0 & 0 & b_{33} & 0\\ 
0 & b_{42} & 0 & 0\\ 
\end{array}\right)\,. \label{extension-Z6Z2-5}
\end{equation}
This solution means that the $\Z_6\rtimes\Z_2$ invariant potential should remain unchanged under the exchange of $\phi_2$ and $\phi_4$ up to phase factors.

We first consider the split extension, which requires that $b^2 = i^q\mathbf{1}_4$. 
Parametrizing the elements of $b$ by phase factors, determining them from the invariance of the potential Eq.~\eqref{V-Z6-2}, 
and imposing $\det b = 1$, $b^\dagger b = \mathbf{1}_4$, we obtain $b$ as
\begin{equation}
b = i^c \left( \begin{array}{cccc}
1 & 0 & 0 & 0\\ 
0 & 0 & 0 & 1\\ 
0 & 0 & \sigma & 0\\ 
0 & 1 & 0 & 0\\ 
\end{array}\right)\,, \label{extension-Z6Z2-6}
\end{equation}
where $c$ is half-integer for $\sigma = 1$ and integer for $\sigma = -1$. 
Therefore, the $D_6$ invariant potential is given by Eq.~\eqref{V-Z6-2} subject to the extra condition
\begin{equation}
\lambda_3 = \sigma \lambda_5\,.\label{extension-Z6Z2-7}
\end{equation}

We also tried constructing the non-split extension $Q_6$ by imposing $b^2 = a_2^3$. 
This could only be done by setting $\lambda_1 = \lambda_2 = 0$. But then the potential
acquires the accidental continuous symmetry given by the arbitrary phase rotations of the first doublet.
Thus, no non-split extensions are possible for $\Z_6$.

\subsection{Extending $\Z_5$}

From \cite{Ivanov:2011ae}, we know that the generator of $\Z_5$ can be chosen in the following way:
\begin{equation}
a = \eta \cdot \diag(\eta,\, \eta^2,\, \eta^3,\, 1) = \diag(\eta^2,\, \eta^{-2},\, \eta^{-1},\, \eta)\,, \quad \eta \equiv e^{2 \pi i/5}\,, \quad \eta^5 = 1.
\label{Z5-a1}
\end{equation}
We verified with the code {\tt 4HDM Toolbox} that all $\Z_5$ invariant 4HDMs can be brought to this choice. 
In addition to the rephasing-insensitive terms $V_0$ given in Eq.~\eqref{V0-general}, 
the Higgs potential invariant under $\Z_5$ contains six terms:
\begin{equation}
\begin{aligned}
V({\Z_5}) & = \lambda_{1} \fdf{2}{1} \fdf{4}{1} + \lambda_{2} \fdf{3}{4}\fdf{2}{4}  + \lambda_{3} \fdf{1}{2} \fdf{3}{2} + \lambda_{4} \fdf{4}{3}\fdf{1}{3}  \\
& +  \lambda_{5} \fdf{1}{3} \fdf{2}{4}  + \lambda_{6} \fdf{4}{1} \fdf{3}{2} +  h.c.\\
\end{aligned}
\label{VZ5}
\end{equation}
The first line here contains the four terms which are linked by the cyclic permutation 
\begin{equation}
\phi_1 \mapsto \phi_4 \mapsto \phi_2 \mapsto \phi_3 \mapsto \phi_1\,.\label{Z5-b-0}
\end{equation}
The second line contains the two possible terms involving all four doublets; these two terms transform into one another upon the same cyclic permutation.
The usefulness of such grouping will become clear once we study the automorphisms of $\Z_5$.
All the coefficients $\lambda_i$ can be complex and, in general, cannot be simultaneously set real by a basis change. 
Thus, the $\Z_5$-invariant 4HDM can be either explicitly $CP$ conserving or $CP$ violating.

The automorphism group for $\Z_5$ is $\Aut(\Z_5) \simeq \Z_4$. The generator $b$ of this automorphism group sends $a$ to $a^2$ or to $a^3$;
the two choices are equivalent.
There is also another automorphism $c$, of order 2, which acts as $a \mapsto a^4 = a^{-1}$.
This automorphism generates the $\Z_2$ subgroup of $\Z_4$; we immediately recognize that $c = b^2$.
Thus, when constructing extensions of $\Z_5$, we can extend it by $\Z_4$ or by $\Z_2$.

\subsubsection{Extension $\Z_5 \rtimes \Z_4$}

We begin with the extension 
$$
\Z_5\rtimes\Z_4 = \langle a, b\, |\, a^5 = e, b^4=e, b^{-1}a b = a^2\rangle\,.\label{extension-Z5Z4-1}
$$
This group of order 20 is also known as $GA(1,5)$, the general affine group over the finite field $\mathbb{F}_5$,
and it has the {\tt GAP Id [20,3]}.
The relation $b^{-1}a b = a^2$ leads to the matrix equation for $b$:
\begin{equation}
ab = ba^{2}\cdot i^r\label{Z5-b-1}\,.
\end{equation}
Since $i = \eta^{5/4}$, we can rewrite this equation as comparison of the powers of $\eta$:
\begin{equation}
\left( \begin{array}{cccc} 
2 & 2 & 2 & 2\\ 
-2 & -2 & -2 & -2\\ 
-1 & -1 & -1 & -1\\ 
1 & 1 & 1 & 1\\ 
\end{array}\right)
\simeq 
\left( \begin{array}{cccc} 
-1 & 1 & -2 & 2\\ 
-1 & 1 & -2 & 2\\ 
-1 & 1 & -2 & 2\\ 
-1 & 1 & -2 & 2\\ 
\end{array}\right)\ + \frac{5}{4}\,r \ \mod 5\,.
\label{extension-Z5Z4-2}
\end{equation}
A solution for the matrix $b$ exists only for $r=0$, 
\begin{equation}
b = \left( \begin{array}{cccc} 
0 & 0 & 0 & b_{14}\\ 
0 & 0 & b_{23} & 0\\ 
b_{31} & 0 & 0 & 0\\ 
0 & b_{42} & 0 & 0\\ 
\end{array}\right)\,. \label{extension-Z5Z4-3}
\end{equation}
We find that $b$ realizes the cyclic permutation given in Eq.~\eqref{Z5-b-0}. 
Each non-zero entry here is a pure phase factor, and
since $b^4 = b_{14}b_{23}b_{31}b_{42}\, \mathbf{1}_4$, we conclude that non-split extensions are not possible. 
By performing an appropriate phase shift basis change, we can set all non-trivial entries 
to be equal to the same phase factor $\exp(i\alpha)$, see Appendix~\ref{appendix-trick}.
We take it out as a universal prefactor and determine $\alpha$ from the condition $\det b = 1$.
In this way we arrive at the extension
\begin{equation}
\Z_5\rtimes \Z_4 \simeq GA(1,5)\,, \quad \mbox{generated by\ }
a = \left( \begin{array}{cccc} 
\eta^2 & 0 & 0 & 0\\ 
0 & \eta^{-2} & 0 & 0\\ 
0 & 0 & \eta^{-1} & 0\\ 
0 & 0 & 0 & \eta\\ 
\end{array}\right)\,,
\quad
b = i^{1/2}\left( \begin{array}{cccc} 
0 & 0 & 0 & 1\\ 
0 & 0 & 1 & 0\\ 
1 & 0 & 0 & 0\\ 
0 & 1 & 0 & 0\\ 
\end{array}\right)\,.
\label{extension-Z5Z4-4}
\end{equation}
In order for the potential $V(\Z_5)$ to be invariant under this group, we must require
\begin{equation}
        \lambda_1  = \lambda_2 = \lambda_3 = \lambda_4\,,\quad \lambda_5  = \lambda_6\,.\label{extension-Z5Z4-5}
\end{equation}
These coefficients can still remain complex.
In addition, we require that the rephasing-insensitive part of the potential $V_0$ is invariant
under Eq.~\eqref{V0-4-0}. Notice that although the potential $V_0$ constrained by Eq.~\eqref{V0-4-0}
possesses accidental symmetries such as $\phi_1 \leftrightarrow \phi_4$, $\phi_2 \leftrightarrow \phi_3$,
this symmetry is not shared by the potential $V(\Z_5)$ even when constrained by Eq.~\eqref{extension-Z5Z4-5}.

As the full scalar potential of the $GA(1,5)$-symmetric 4HDM contains only 10 real free parameters,
one can expect many correlations among scalar properties. 
We delegate the phenomenological analysis of this model to a future paper.
We are not aware of any NHDM study which uses the group $GA(1,5)$.

\subsubsection{Extension $\Z_5 \rtimes \Z_2$}

Let us now construct the second extension:
$$
\Z_5\rtimes\Z_2\simeq D_{5} = \langle a, c\, |\, a^5 = e, c^2=e, c^{-1}a c = a^{-1}\rangle\,.\label{Z5-c-1}
$$
Instead of relying on $b^2$, we write and solve the matrix equation on $c$ and arrive at the solution
\begin{equation}
c = \left( \begin{array}{cccc} 
0 & c_{12} & 0 & 0\\ 
c_{21} & 0 & 0 & 0\\ 
0 & 0 & 0 & c_{34}\\ 
0 & 0 & c_{43} & 0\\ 
\end{array}\right)\,, \label{extension-Z5Z2-1}
\end{equation}
which is of the same form as $b$ in Eq.~\eqref{extension-Z6Z2-2}. All the considerations given there apply to this case,
and we arrive at the 4HDM invariant under 
\begin{equation}
\Z_5\rtimes\Z_2\simeq D_{5}\,, \quad \mbox{generated by\ }
a = \left( \begin{array}{cccc} 
\eta^2 & 0 & 0 & 0\\ 
0 & \eta^{-2} & 0 & 0\\ 
0 & 0 & \eta^{-1} & 0\\ 
0 & 0 & 0 & \eta\\ 
\end{array}\right)\,,
\quad
c = \left( \begin{array}{cccc} 
0 & 1 & 0 & 0\\ 
1 & 0 & 0 & 0\\ 
0 & 0 & 0 & 1\\ 
0 & 0 & 1 & 0\\ 
\end{array}\right)\,. \label{extension-Z5Z2-2}
\end{equation}
This symmetry arises if the conditions in Eq.~\eqref{V0-2-2} on $V_0$ are fulfilled and the parameters $\lambda$ in Eq.~\eqref{VZ5} satisfy
\begin{equation}
\lambda_1 = \lambda_3\,, \quad  
\lambda_2 = \lambda_4\,,\label{extension-Z5Z2-3}
\end{equation}
while $\lambda_5$ and $\lambda_6$ are unconstrained.
The rephasing-insensitive potential $V_0$ must satisfy the conditions Eq.~\eqref{V0-2-2}.

\subsection{Extending $\Z_4$}\label{subsection-Z4}

\subsubsection{The three versions of $\Z_4$ in the 4HDM}

In the previous subsections, we considered the groups $\Z_n$, $n = 5,6,7,8$, which were not available in the 3HDM scalar sector \cite{Ivanov:2011ae}.
The case of $\Z_4$ is different. It was already present in the 3HDM study, where it was the largest cyclic symmetry group 
realizable in the 3HDM scalar sector, with the generator $\diag(\eta, \eta^{-1}, 1)$, where $\eta \equiv \exp(2\pi i/4)$.
Since $\eta = i$, we prefer not abuse the notation and will replace $\eta$'s with $i$'s throughout this subsection.

When embedding it in the 4HDM scalar sector, one could just take this generator and assume that the fourth doublet transforms under $\Z_4$
by multiplication of some power of $i$: $1$, $-1$, or $i$ (with the choice $-i$ being equivalent to $i$ upon $\phi_1\leftrightarrow \phi_2$). 
This construction leads us to three nonequivalent $\Z_4$ symmetries in the 4HDM, which we present in a slightly rearranged form:
\begin{eqnarray}
\mbox{$\Z_4$ option 1:} && a_1 = \sqrt{i}\cdot\diag(i,-1, -i, 1)\label{extension-Z4-a1}\\
\mbox{$\Z_4$ option 2:} && a_2 = \diag(i,-i, 1, 1)\label{extension-Z4-a2}\\
\mbox{$\Z_4$ option 3:} && a_3 = i^{3/4}\,\diag(i,i,-i, 1)\label{extension-Z4-a3}
\end{eqnarray}
The powers of $i$ are introduced to ensure that $\det a = 1$.
Option 1 can be called the ``fully represented $\Z_4$'' because the four doublets transform
as the four possible singlets of $\Z_4$.
Options 2 and 3 contain 2D invariant subspaces. 

Any other choice of the $\Z_4$ generator either can be reduced to one of those or will lead
to a continuous rephasing symmetry. For example, if one chooses
the generator $\sqrt{i}\cdot \diag(i,i,1,1)$ with two 2D invariant subspaces and tries to construct
a potential invariant under it, one will unavoidably obtain the accidental symmetry
$(e^{i\alpha}, e^{i\alpha}, e^{-i\alpha}, e^{-i\alpha})$.

The three options for the $\Z_4$ generator represent truly distinct cases, which cannot be linked by any basis change. 
Indeed, $a_1$ does not contain any 2D invariant subspace, while $a_3$ differs from $a_2$
by the fact that $a_3^2$ has a 3D invariant subspace.
This situation is reminiscent of two nonequivalent $U(1)$ groups which exist already in the 3HDM \cite{Ivanov:2011ae}.
With the aid of the code {\tt 4HDM Toolbox}, 
we verified that, in all the cases of the symmetry group $\Z_4$, we obtain the generator of the form
Eq.~\eqref{extension-Z4-a1}, Eq.~\eqref{extension-Z4-a2}, or Eq.~\eqref{extension-Z4-a3}, up to permutations and rephasing.

Below, we will extend all three versions of the $\Z_4$ group. Since $\Aut(\Z_4) \simeq \Z_2$, we can have
only two possibilities for non-abelian extensions: the split extension $\Z_4\rtimes \Z_2 \simeq D_4$ and the non-split extension
$\Z_4\,.\,\Z_2 \simeq Q_4$, see Section~\ref{subsection-split}. Thus, when constructing the generator $b$ of the $\Z_2$ group,
we will, in principle, need to check two options: $b^2 = e$ or $b^2 = a^2$, for each version of the $\Z_4$ group.

\subsubsection{Extending the fully represented $\Z_4$}\label{subsubsection-fully-rep-Z4}

The potential invariant under the generator $a_1$ in Eq.~\eqref{extension-Z4-a1} contains, apart from the rephasing invariant $V_0$,
the following terms:
\begin{equation}
\begin{aligned}
V_1(\Z_4) & = \lambda_1\fdf{1}{2}\fdf{1}{4} + \lambda_2\fdf{2}{1}\fdf{2}{3} + \lambda_3\fdf{1}{3}^2 \\ 
& + \lambda_4\fdf{4}{1}\fdf{4}{3} + \lambda_5\fdf{3}{2}\fdf{3}{4} + \lambda_6\fdf{2}{4}^2 \\
& + \lambda_7\fdf{1}{2}\fdf{4}{3} + \lambda_8\fdf{1}{3}\fdf{4}{2} + \lambda_9\fdf{1}{3}\fdf{2}{4} + \lambda_{10}\fdf{1}{4}\fdf{2}{3} + h.c.
\end{aligned}
\label{V1-Z4}
\end{equation}
These terms can be found by writing the $\Z_4$ charges of all possible quadratic monomials $\fdf{i}{j}$
and multiplying the terms with opposite charges to obtain the $\Z_4$-invariant terms.
As this potential has ten complex free parameters, it is in general impossible to set them simultaneously real
by any basis change; thus, the $\Z_4$ 4HDM model can be either explicitly $CP$ conserving or $CP$ violating. 

The automorphism group of $\Z_4$ is $\Z_2$, whose generator $b$ sends $a_1$ to $a_1^3$. 
Thus, we have matrix equation for the generator $b$ of $\Z_2$,
\begin{equation}
a_1b = ba_1^3\cdot i^r\,,
\end{equation}
which can be represented as the equation for the powers of $i$:
\begin{equation}\frac{1}{2} + 
\left( \begin{array}{cccc} 
1 & 1 & 1 & 1\\ 
2 & 2 & 2 & 2\\ 
3 & 3 & 3 & 3\\ 
0 & 0 & 0 & 0\\ 
\end{array}\right)
\simeq \frac{3}{2} + 
\left( \begin{array}{cccc} 
3 & 2 & 1 & 0\\ 
3 & 2 & 1 & 0\\ 
3 & 2 & 1 & 0\\ 
3 & 2 & 1 & 0\\ 
\end{array}\right)\ + \,r \ \mod 4\,.
\label{extension-Z4Z2-2}
\end{equation}
This set of matching conditions has one solution for each $r = 0,1,2,3$.
However solutions for $r = 0$ and $2$ are mapped onto each other by renaming the doublets,
and so are solutions for $r = 1$ and $3$. Thus, we have two non-equivalent classes of $b$,
which, using the rephasing basis change freedom, can be represented as
\begin{equation}
b = 
\begin{pmatrix}
0 & e^{i\alpha} & 0 & 0 \\
e^{i\alpha} & 0 & 0 & 0 \\
0 & 0 & 0 & e^{i\beta} \\
0 & 0 & e^{i\beta} & 0 
\end{pmatrix}
\,,\quad 
b' = 
\begin{pmatrix}
0 & 0 & e^{i\alpha} & 0 \\
0 & e^{i\gamma_2} & 0 & 0 \\
e^{i\alpha} & 0 & 0 & 0 \\
0 & 0 & 0 & e^{i\gamma_4} 
\end{pmatrix}
\label{extension-Z4Z2-bb}
\end{equation}
Let us first consider $b$. Squaring it leads to $b^2 = \diag(x,x,y,y)$, which cannot match any nontrivial power of $a_1$,
which forbids the non-split extension. 
Next, by requiring $\det b = 1$ and $b^2 = i^{q}\mathbf{1}_4$, and making use of rephasing basis change freedom,
we obtain the unique expression for $b$, up to the omnipresent powers of $i$:
\begin{equation}
b = 
\begin{pmatrix}
0 & 1 & 0 & 0 \\
1 & 0 & 0 & 0 \\
0 & 0 & 0 & 1 \\
0 & 0 & 1 & 0 
\end{pmatrix}\,.
\label{extension-Z4Z2-b2}
\end{equation}
Thus, we obtain the group $D_4$ generated by $a_1$ in Eq.~\eqref{extension-Z4-a1}
and by $b$ in Eq.~\eqref{extension-Z4Z2-b2}.
Requiring that the potential $V_1(\Z_4)$ be invariant under $b$ leads to the following relations:
\begin{equation}
\lambda_1 = \lambda_2\,,\quad \lambda_3 = \lambda_6\,,\quad \lambda_4 = \lambda_5\,,\quad \lambda_7,\lambda_8\in\mathbb{R}\,.
\label{extension-Z4Z2-b3p}
\end{equation}
The number of real parameters is reduced from 20 in Eq.~\eqref{V1-Z4} to 12.
The potential still contains complex parameters, so that the model allows for explicit $CP$ violation.

Next, we consider $b'$ in Eq.~\eqref{extension-Z4Z2-bb} and still aim to construct 
the split extension $\Z_4 \rtimes \Z_2 \simeq D_4$. Repeating the above analysis, we get 
the following generic expression for $b'$ which covers all sign choices:
\begin{equation}
b' = i^c
\begin{pmatrix}
0 & 0 & 1 & 0 \\
0 & \sigma & 0 & 0 \\
1 & 0 & 0 & 0 \\
0 & 0 & 0 & 1 
\end{pmatrix}\,.
\label{extension-Z4Z2-b3}
\end{equation}
Here, $\sigma = \pm 1$ and $c$ is integer for $\sigma = - 1$ and half-integer for $\sigma = +1$. 
Thus, the two transformations $a_1$ and $b'$ generate another version of the $D_4$-invariant 4HDM.
In order for $V_1(\Z_4)$ to be invariant under $b'$, its parameters must satisfy
\begin{equation}
\lambda_5 = \sigma \lambda_1\,, \quad 
\lambda_{10} = \sigma \lambda_7^*\,, \quad 
\lambda_9 = \sigma \lambda_8^*\,, \quad 
\lambda_3 \in \mathbb{R} \,.\label{extension-Z4Z2-b4}
\end{equation}
We are left with 13 real free parameters in Eq.~\eqref{V1-Z4}.

In addition, we can use $b'$ from Eq.~\eqref{extension-Z4Z2-bb} to construct the non-split extension
$\Z_4\,.\,\Z_2 \simeq Q_4$; this is the first example in which we encounter this possibility. 
Let us denote this version of $b'$ as $b''$.
Then, $(b'')^2 = \diag(x,y,x,z)$ and it can match $a_1^2 = \diag(-i, i, -i, i)$.
Combining this matching with $\det b'' = 1$, we obtain the following general expression for $b''$:
\begin{equation}
b'' = i^c
\begin{pmatrix}
0 & 0 & 1 & 0 \\
0 & -i\sigma & 0 & 0 \\
1 & 0 & 0 & 0 \\
0 & 0 & 0 & i 
\end{pmatrix}\,,
\label{extension-Z4Z2-b5}
\end{equation}
with the same convention for the power $c$ as stated below Eq.~\eqref{extension-Z4Z2-b3}.
Imposing $b''$ symmetry on the potential leads to the same type of restrictions as in Eq.~\eqref{extension-Z4Z2-b4}
and, in addition, eliminates the $\lambda_2$ and $\lambda_4$ terms. Thus, we are left with 9 real free parameters
in the rephasing sensitive part of the $Q_4$-invariant 4HDM.

\subsubsection{Extending $\Z_4$ option 2}

Next we turn to the $\Z_4$ symmetry group generated by $a_2$ defined in Eq.~\eqref{extension-Z4-a2}.
The rephasing-sensitive part of the potential invariant under this generator is
\begin{equation}
\begin{aligned}
V_2(\Z_4) & = m_{34}^2\fdf{3}{4} + \lambda_1\fdf{1}{2}^2 + \lambda_2\fdf{3}{4}^2
+ \lambda_3\fdf{3}{1}\fdf{3}{2} + \lambda_4\fdf{4}{1}\fdf{4}{2} \\
 & + \lambda_5\fdf{1}{3}\fdf{2}{4} + \lambda_6 \fdf{1}{4}\fdf{2}{3} + h.c.
\end{aligned}
\label{V2-Z4}
\end{equation}
This potential contains seven complex free parameters;
notice the presence of a new quadratic term. 

To extend this $\Z_4$ by a $\Z_2$ generated by a new transformation $c$, we follow 
the same strategy as before and obtain the following generic solution:
\begin{equation}
c = 
\begin{pmatrix}
0 & c_{12} & 0 & 0 \\
c_{21} & 0 & 0 & 0 \\
0 & 0 & c_{33} & c_{34} \\
0 & 0 & c_{43} & c_{44} \\
\end{pmatrix}\,.\label{extension-Z4-c1}
\end{equation}
The presence of the $2\times 2$ block in the $(\phi_3,\phi_4)$ subspace is a clear consequence of the fact that this subspace
is invariant under $\Z_4$.
However, $c^2$ must be diagonal: proportional either to $\mathbf{1}_4$ for the split extension $D_4$ 
or to $a_2^2$ for the non-split extension $Q_4$.
This leaves us with two possible shapes of this $2\times2$ block:
\begin{equation}
\mmatrix{c_{33}}{0}{0}{c_{44}}\quad \mbox{or}\quad \mmatrix{0}{c_{34}}{c_{43}}{0}\,.\label{extension-Z4-c2}
\end{equation}
Repeating the above analysis, we found that both options allow for a split and a non-split extension.
For example, the first option leads to the split extension $D_4$ which leaves $\phi_4$ invariant.
Therefore, this is exactly the same $D_4$ as was found in the 3HDM \cite{Ivanov:2012fp}.
For the non-split extension, we get
\begin{equation}
c = 
i^q\begin{pmatrix}
0 & i & 0 & 0 \\
i & 0 & 0 & 0 \\
0 & 0 & \sigma & 0 \\
0 & 0 & 0 & 1 \\
\end{pmatrix}\,,\label{extension-Z4-c3}
\end{equation}
where $\sigma = \pm 1$, with the corresponding value of $q$ (integer for $\sigma = + 1$ and half-integer for $\sigma = -1$).
Imposing this symmetry on $V_{2}(\Z_4)$ eliminates the $\lambda_3$ and $\lambda_4$ terms, and the rephasing sensitive potential 
can then be compactly written as
\begin{equation}
V_2(Q_4) = \delta_{\sigma,1}m_{34}^2 \fdf{3}{4} + \lambda_1\fdf{1}{2}^2 + \lambda_2\fdf{3}{4}^2
+ \lambda_5\left[ \fdf{1}{3}\fdf{2}{4} -\sigma \fdf{1}{4}\fdf{2}{3}\right] + h.c.
\label{V2-Q4}
\end{equation}
with a real $\lambda_1$ and the quadratic term present only for $\sigma = +1$.
Notice the crucial role of $\phi_4$: although it is a $Q_4$ singlet, its presence allows us to build terms which
were absent in the 3HDM. As a result, we now have a realizable $Q_4$-invariant 4HDM without any accidental continuous symmetry,
the situation found in \cite{Ivanov:2012fp} impossible in the 3HDM.
The second option in Eq.~\eqref{extension-Z4-c2}, too, leads to viable split and non-split extensions.

\subsubsection{Extending $\Z_4$ option 3}

Finally, we consider the $\Z_4$ symmetry group generated by $a_3$ defined in Eq.~\eqref{extension-Z4-a3}.
The rephasing-sensitive part of the potential is
\begin{eqnarray}
V_3(\Z_4) &= & m_{12}^2\fdf{1}{2} + \lambda_1\fdf{1}{2}^2 + \lambda_2\fdf{1}{3}^2 + \lambda_3\fdf{2}{3}^2\nonumber\\
&& + \lambda_4\fdf{1}{4}\fdf{3}{4} + \lambda_5\fdf{2}{4}\fdf{3}{4} + \lambda_6\fdf{1}{3}\fdf{2}{3} + h.c.
\label{V3-Z4}
\end{eqnarray}
This potential contains seven complex free parameters, including one in the quadratic part. 

In order to extend this $\Z_4$, we need to solve $a_3b = ba_3^3\cdot i^r$ for $b$.
However, the prefactor $i^{3/4}$ prevents us from finding a non-trivial solution, in a way similar to our $\Z_8$ analysis.
Thus, we arrive at a peculiar $\Z_4$-invariant 4HDM which does not admit any extension.

To summarize our discussion of possible extensions of the symmetry group $\Z_4$ available in the 4HDM, we found that 
there are two non-equivalent $\Z_4$ symmetry groups in the 4HDM which can be extended both to $D_4 \simeq \Z_4\rtimes \Z_2$ 
and to $Q_4 \simeq \Z_4\,.\,\Z_2$ in a variety of non-equivalent ways.
We explicitly constructed the generators of these groups and showed how the rephasing-sensitive part of the potential
is shaped by them.
We stress that the quaternion symmetry group $Q_4$ is a novel option for the 4HDM model building, 
which was unavailable within the 3HDM.
In addition, there exists a peculiar realization of $\Z_4$ in the 4HDM which does not admit any extension by $\Aut(\Z_4)$.

\subsection{Extending $\Z_3$}\label{subsection-Z3}

Just as for the case of $\Z_4$, the symmetry group $\Z_3$ was also realizable in the 3HDM \cite{Ivanov:2011ae}.
Its generator in the 3HDM can be written as $\diag(1, \eta, \eta^{-1})$, where $\eta \equiv \exp(2\pi i/3)$, which is often denoted $\omega$.
Since $\eta^3 = 1$, this generator can be written also as $\diag(1, \eta, \eta^2)$. This choice of $\Z_3$ charges is unique up to permutation;
if one tries to build a 3HDM based on, say, $\diag(1, 1, \eta)$, one would end up with a potential invariant under the continuous $U(1)$ symmetry.

Within the $\Z_3$ symmetric 4HDM, we again need to assign all three $\Z_3$ charges to the four Higgs doublets, 
and one of the charges will be used twice. As a result, we have only one possibility, up to permutations and the overall $\Z_3$ charge shifts.
The generator can be written as
\begin{equation}
a = \diag(\eta,\, \eta^2,\,1,\, 1)\,, \quad \eta \equiv e^{2 \pi i/3}\,, \quad \eta^3 = 1.
\label{Z3-a}
\end{equation}
The $\Z_3$-invariant 4HDM potential contains, in addition to $V_0$ given in Eq.~\eqref{V0-general}, the following terms:
\begin{eqnarray}
V_1(\Z_3) &= & m_{34}^2\fdf{3}{4} + \lambda_1\fdf{3}{4}^2 
+ \lambda_2\fdf{1}{2}\fdf{1}{3} + \lambda_3\fdf{2}{3}\fdf{2}{1} \\
&& + \lambda_4\fdf{1}{2}\fdf{1}{4} + \lambda_5\fdf{2}{4}\fdf{2}{1} 
+ \lambda_6\fdf{3}{1}\fdf{3}{2} \nonumber\\
&&+ \lambda_7\fdf{4}{1}\fdf{4}{2} 
+ \lambda_8\fdf{3}{1}\fdf{4}{2} + \lambda_9\fdf{4}{1}\fdf{3}{2}+ h.c.
\label{V1-Z3}
\end{eqnarray}
All of these ten coefficients can be complex.

The automorphism group of $\Z_3$ is $\Z_2$. 
Group-theoretically, we get only one non-abelian extension:
$\Z_3 \rtimes \Z_2 \simeq S_3$. The non-trivial automorphism is given by
\begin{equation}
b = 
i^q\begin{pmatrix}
0 & 1 & 0 & 0 \\
1 & 0 & 0 & 0 \\
0 & 0 & \sigma & 0 \\
0 & 0 & 0 & 1 \\
\end{pmatrix}\,,\label{extension-S3-b}
\end{equation}
where $\sigma = \pm 1$ and $q$ is half-integer for $\sigma = + 1$ and integer for $\sigma = -1$.
The version with $\sigma = -1$ can be called the fully-represented $S_3$ because in this case 
the irreducible representation decomposition of the four Higgs doublets is $2 + 1 + 1'$.

The two versions of the generator $b$ constrain the potential $V_1(Z_3)$ in Eq.~\eqref{V1-Z3}
in a slightly different manner.
In both cases, we have 
\begin{equation}
\lambda_2 = \sigma \lambda_3\,, \quad \lambda_4 = \lambda_5\,, \quad \lambda_8 = \sigma\lambda_9\,.\label{extension-S3-1}
\end{equation}
In addition, the fully represented $S_3$ model forbids the non-trivial quadratic term: $m_{34}^2 = 0$.
Also, in both cases, the rephasing-insensitive potential $V_0$ given in Eq.~\eqref{V0-general}
satisfies the conditions of the type of Eq.~\eqref{V0-2-1-1} but written for indices $3,4$ instead of $1,2$.

Let us finally mention that the symmetry group $\Z_2$ does not have non-trivial automorphisms and cannot be extended.
We only remark that there exist two inequivalent implementations of $\Z_2$ in the 4HDM corresponding to the 
generators $\sqrt{i}\diag(1,1,1,-1)$ and $\diag(1,1,-1,-1)$. They lead to different constraints on the scalar potential,
which are straightforward to write down.


\section{Discussion and conclusions}\label{section-conclusions}

\begin{table}[H]
\centering
\begin{tabular}[t]{cccccccc}
\toprule
$A$ & extension & $G$ & $|G|$ & irreps & conditions on $V_0$ & comments \\
\midrule
$\Z_2$ & --- & --- & --- & --- & --- & --- \\[1mm]
$\Z_3$ & $\Z_3\rtimes \Z_2$ & $S_3$ & 6 & $1+1+2$ & Eq.~\eqref{V0-2-1-1} & Section~\ref{subsection-Z3} \\[1mm]
$\Z_4$ & $\Z_4\rtimes \Z_2$ & $D_4$ & 8 & \begin{tabular}[t]{@{}c@{}}$1+1+2$\\or $2+2$\end{tabular} 
& \begin{tabular}[t]{@{}c@{}} Eq.~\eqref{V0-2-1-1} \\ or Eq.~\eqref{V0-2-2}  \end{tabular}    & Section~\ref{subsection-Z4} \\
	   & $\Z_4\,.\, \Z_2$ & $Q_4$ & 8 & $1+1+2$ & Eq.~\eqref{V0-2-1-1} & Section~\ref{subsection-Z4} \\[1mm]
$\Z_5$ & $\Z_5\rtimes \Z_4$ & $GA(1,5)$ & 20 & $4$ & Eq.~\eqref{V0-4-0} & Eqs.~Eq.~\eqref{extension-Z5Z4-4}, Eq.~\eqref{extension-Z5Z4-5} \\
	   & $\Z_5\rtimes \Z_2$ & $D_5$ & 10 & $2+2$ & Eq.~\eqref{V0-2-2} & Eqs.~Eq.~\eqref{extension-Z5Z2-2}, Eq.~\eqref{extension-Z5Z2-3} \\[1mm]
$\Z_6$ & $\Z_6\rtimes \Z_2$ & $D_6$ & 12 & \begin{tabular}[t]{@{}c@{}}$2+2$\\or $1+1+2$\end{tabular} & 
\begin{tabular}[t]{@{}c@{}} Eq.~\eqref{V0-2-2} \\ or Eq.~\eqref{V0-2-1-1}  \end{tabular}   
& \begin{tabular}[t]{@{}c@{}} Eqs.~Eq.~\eqref{extension-Z6Z2-3}, Eq.~\eqref{VZ6Z2-1} \\ or Eqs.~Eq.~\eqref{extension-Z6Z2-6}, Eq.~\eqref{extension-Z6Z2-7}   \end{tabular} \\[1mm]
$\Z_7$ & $\Z_7\rtimes \Z_3$ & $T_7$ & 21 & $1+3$ & Eq.~\eqref{V0-3-1}& Eqs.~Eq.~\eqref{extension-Z7Z3-4}, Eq.~\eqref{VT7-1}  \\[1mm]
$\Z_8$ & --- & --- & ---  & --- \\
\bottomrule
\end{tabular}
\caption{The summary table of non-abelian groups $G$ available for the 4HDM scalar sector
constructed by extension of a cyclic group by its automorphisms and
not leading to accidental symmetries.}
\label{table-non-abelian}
\end{table}

Table~\ref{table-non-abelian} summarizes the results of this work.
Starting from the full list of cyclic symmetry groups $A$ available in the 4HDM scalar sector, which we know from
\cite{Ivanov:2011ae}, we found, for each $A$, its automorphism group and 
built all possible non-abelian extensions of $A$ by its automorphisms which fit in the 4HDM scalar sector.
The list of non-abelian symmetry groups which we obtain in this way is the following:
\begin{equation}
\mbox{$G$ in 4HDM:}\qquad S_3\,,\quad 
D_4\,,\quad 
Q_4\,,\quad 
GA(1,5)\,,\quad 
D_5\,,\quad 
D_6\,,\quad T_7\,.
\label{4HDM-non-abelian}
\end{equation}
For each of these groups, we give in Table~\ref{table-non-abelian} the order of the group 
and how the four Higgs doublets are decomposed into the irreducible representations (irreps) of the group.
Notice that in certain cases more than one choice of irrep decomposition is available.
Notice also that, when indicating the irreps, we only give their dimensions without distinguishing irreps 
of the same dimension such as $1$ or $1'$; the exact assignment can be understood from the main text.
We also indicate in the table which set of conditions on the parameters of $V_0$, given in Eq.~\eqref{V0-general},
must be applied, as well as the explicit forms of the generators and the constraints on the coefficients
of the rephasing sensitive part of the potential.
This Table gives the full list of non-abelian Higgs family symmetry groups which one can construct in the 4HDM scalar sector
as extensions of cyclic groups $A$.

This study brought up several issues which we believe may be not so well known by the community.
\begin{itemize}
\item
If we restrict the field content, there is only a limited choice of cyclic groups which can be used for model building.
For example, within the scalar sector of the 4HDM, we can only use $\Z_n$ up to $n=8$.
Larger cyclic groups unavoidably lead to an accidental continuous symmetry.
\item
For a given cyclic group $\Z_n$, not all $\Z_n$ charge assignments are available.
For example, when building a $\Z_8$-invariant 4HDM, we have a unique charge assignment, up to permutations and conjugations.
Trying to build another $\Z_8$-invariant 4HDM with a different assignment of charges will again lead to an accidental continuous symmetry.
\item
On the other hand, this uniqueness feature does not apply to all cyclic groups.
In particular, we demonstrated that there are 
two inequivalent realizations of the symmetry group $\Z_6$ and
three inequivalent realizations of $\Z_4$,
leading to models with distinct number of free parameters and different options for non-abelian extensions.
\item
Imposing certain cyclic groups automatically leads to explicit $CP$ conservation.
\item
Although the group $\Z_4$ was already available in the 3HDM, it offers several new features within the 4HDM:
three inequivalent realizations, the possibility of $CP$ violation, and the possibility of a non-split extension
$Q_4$.
\end{itemize}
The list Eq.~\eqref{4HDM-non-abelian} does not exhaust all finite non-abelian discrete groups available in 4HDM.
One can also take other abelian groups $A$, which are products of cyclic groups available in the 4HDM:
$\Z_2\times \Z_2$, $\Z_4\times \Z_2$, $\Z_2\times \Z_2\times \Z_2$, and $\Z_4\times \Z_4$.
Their automorphism groups were given in Table~\ref{table-abelian}, right, and we expect them
to produce several new non-abelian extensions. This task turns out to be more laborious and is delegated to a follow-up paper.

Moreover, the key argument using which Refs.~\cite{Ivanov:2012fp,Ivanov:2012ry} could exhaust all 
finite non-abelian groups within the 3HDM scalar sector no longer applies to the 4HDM case.
As a result, there is no guarantee that all non-abelian groups in the 4HDM can be constructed 
as extensions of abelian groups. Additional group-theoretical insights are needed
to complete the classification of finite symmetry groups in the 4HDM.

To summarize, this paper presents the first step of a systematic study of 
finite non-abelian symmetry groups available for the 4HDM scalar sector.
We borrowed from the 3HDM --- and further developed --- the algorithmic strategy 
of constructing finite non-abelian groups 
as extensions of abelian groups by their automorphisms.
We applied this strategy to all cyclic groups available for the 4HDM
and obtained several 4HDM models based on non-abelian groups.
The same strategy can also be applied to products of cyclic groups which are also known to exist in the 4HDM;
this work is delegated to a follow-up paper.

Group theory of the 4HDM scalar sector turned out to be more involved than in the 3HDM.
We suspect that the above strategy does not cover all the non-abelian groups available for the 4HDM scalar sector,
and additional insights may be needed to complete the classification.
Nevertheless, even within the class of symmetry-based 4HDM models we have obtained,
there are interesting cases which possess sufficiently constrained potentials and may lead to peculiar phenomenology. 
Discovering these symmetry-driven phenomenological features 
and developing observational criteria which would be able to distinguish the underlying symmetry group
of the 4HDM scalar sector just from the properties of the physical scalars is an extensive research program, 
which will be addressed in future works.

\section*{Acknowledgments}

This work was supported by the Fundamental Research Funds for the Central Universities, Sun Yat-sen University,
China.


\appendix

\section{The Smith normal form technique}\label{appendix-SNF}

Consider an arbitrary interaction lagrangian which contains several complex fields.
It may happen that certain global phase rotations leave this lagrangian invariant.
Such a transformation represents a rephasing symmetry of the lagrangian; 
all such symmetries form the rephasing symmetry group.
A question then arises: how can one establish the rephasing symmetry group of a given lagrangian?

There exists a completely general algorithmic procedure which solves this task.
It is based on the so-called Smith normal forms (SNF) and, for any input lagrangian,
gives its rephasing symmetry group and the list of ``charges'' under the corresponding finite or continuous phase rotations. 
The detailed exposition of the technique can be found in \cite{Ivanov:2011ae}, with illustrations
for the 3HDM, 4HDM, and the general NHDM scalar sector.
A somewhat simplified explanation of the general strategy can be also found in section 2 of \cite{Ivanov:2013bka}
with examples from the NHDM Yukawa sector.
Here, we briefly recapitulate the method and describe our Python code {\tt 4HDM Toolbox} 
which makes use of the SNF technique to identify all realizations of cyclic symmetry groups in the 4HDM scalar sector.

To give an example, consider the 4HDM scalar sector and suppose the potential contains $k$ rephasing-sensitive terms
(not counting the complex conjugated ones).
Let us perform phase shifts of the four doublets as $\phi_j \mapsto e^{i\alpha_j}\phi_j$, 
with $\alpha_1, \dots, \alpha_4$ all independent.
Then, the first term of the potential picks up a phase rotation which can be generically written as 
$d_{11}\alpha_1 + d_{12}\alpha_2 + d_{13}\alpha_1 + d_{14}\alpha_4$, where the integer coefficients $d_{1j}$
indicate the power of $\phi_j$ in the first interaction term.
For example, the term $m_{14}^2 \fdf{1}{4}$ corresponds to $d_{1j} = (-1, 0, 0, 1)$,
while the second term $\lambda\fdf{1}{2}\fdf{4}{3}$ corresponds to $d_{2j} = (-1, 1, 1, -1)$.

If we have $k$ rephasing-sensitive terms, we obtain $k$ such rows of coefficients $d_{ij}$ and write them together
as a rectangular integer-valued matrix $D$. For example,
\begin{equation}
m_{14}^2\fdf{1}{4} + \lambda\fdf{1}{2}\fdf{4}{3} + \lambda'\fdf{2}{3}^2\quad \Rightarrow \quad D = 
\begin{pmatrix}
-1 & 0 & 0 & 1 \\
-1 & 1 & 1 & -1 \\
0 & -2 & 2 & 0 
\end{pmatrix}
\label{example-SNF-1}
\end{equation}
The complex conjugated terms differ by the overall minus sign; we drop them because they do not change the subsequent analysis.
Notice that the monomial $\fdf{1}{2}\fdf{4}{3}$ and monomial $\fdf{1}{3}\fdf{4}{2}$ are both represented as $(-1,1,1,-1)$ 
in the matrix, and we do not repeat the same row of coefficients. 

The lagrangian is invariant under some phase rotations if and only if there exist non-trivial solutions $(\alpha_1, \dots, \alpha_N)$
to the following matrix equation:
\begin{equation}
d_{ij}\alpha_j = 2\pi n_i\,, \quad n_i \in \mathbb{Z}\,.\label{SNF}
\end{equation}
As is explained in \cite{Ivanov:2011ae}, there exists a set of simple transformation rules
which can bring the matrix $D$ to its diagonal form and, at the same time, preserves
the set of solutions.
The sequence of these elementary steps can be represented by multiplication of the left and right integer-valued matrices:
\begin{equation}
S\cdot D\cdot T = N\,,\label{example-SNF-2}
\end{equation}
where the diagonal matrix $N$ is called the Smith normal form.
In the above example,
\begin{equation}
S\cdot D\cdot T = 
\begin{pmatrix}
-1 & 1 & 0\\
-1 & 2 & 1 \\
-2 & 2 & 1 
\end{pmatrix}
\begin{pmatrix}
-1 & 0 & 0 & 1 \\
-1 & 1 & 1 & -1 \\
0 & -2 & 2 & 0 
\end{pmatrix}
\begin{pmatrix}
0 & -1 & 4 & 1 \\
1 & 0 & -1 & 1 \\
0 & 0 & 1 & 1 \\
0 & 0 & 0 & 1 \\
\end{pmatrix}
 = 
\begin{pmatrix}
1 & 0 & 0 & 0\\
0 & 1 & 0 & 0\\
0 & 0 & 4 & 0\\
\end{pmatrix}
 = N\,.
 \label{example-SNF-3}
\end{equation}
Once we know $N$ and its diagonal entries $d_i$, the system Eq.~\eqref{SNF} transforms into several uncoupled equations:
\begin{equation}
d_i\tilde \alpha_i = 2\pi \tilde n_i\,, \quad \tilde n_i \in \mathbb{Z}\,,\label{SNF-2}
\end{equation}
with unconstrained $\tilde \alpha_j$ for $j > k$ or for any situation with $d_i = 0$.
The solution to this system is elementary, and it yields the group $\Z_{d_1}\times\Z_{d_2}\times \dots\times \Z_{d_k} \times [U(1)]^{N-k}$.
In the 4HDM, we always have the final $U(1)$ (the common phase rotation), and we are interested
in the rephasing group factored by it.

For example, in the above example Eq.~\eqref{example-SNF-3}, we deduce that the potential contains the symmetry group $\Z_4$.
The absence of zeros on the main diagonal indicates that there is no accidental continuous symmetry.
In addition, the third column of the matrix $T$ stores the $\Z_4$ charges of the four doublets, 
which can be expressed as powers of $i$:
\begin{equation}
a = \diag(i^4,i^{-1},i^1, i^0) = \diag(1, -i, i, 1)\,.
\end{equation}

In the scalar sector of the 4HDM, there are 30 rephasing-sensitive monomials (counted up to conjugation), 
which include 6 quadratic terms and 24 quartic terms. 
Some of these terms transform under rephasing in the same way, such as $\fdf{1}{2}\fdf{3}{4}$ and $\fdf{1}{4}\fdf{3}{2}$.
Thus, we have in total 27 differently transforming monomials.

If we want to exhaust all possible situations with respect to rephasing symmetry transformations,
we would need to pick up all possible combinations of several monomials from this list and compute the corresponding SNF matrix.
This would be a tedious task to do by hand.
In \cite{Ivanov:2011ae}, bypassing such case-by-case checks, a theorem was proved which showed that any
finite abelian symmetry group $A$ with order $|A| \le 8$ can be realized as a rephasing symmetry of the 4HDM scalar sector.
However for specific realizations of each of these $A$, one would still need to verify various cases.

To facilitate this study, we wrote a Python code {\tt 4HDM Toolbox} available at \cite{TheCode} which does it automatically.
The user defines which cyclic symmetry group $\Z_n$ should be searched for,
and the code iteratively checks all combinations of three distinct monomials and finds which cases yield
the desired group $\Z_n$. Then, for these cases, it computes the $\Z_n$ charges and, finally,
completes the potential by adding all terms which are invariant under this particular realization of the group $\Z_n$.
In this way, we could verify, for any $\Z_n$, that we do not miss any specific realization.

\section{Simplifying permutations}\label{appendix-trick}

Suppose that a 4HDM potential is invariant under an abelian group of phase shifts generated by $a$
and an exchange of two doublets such as $\phi_1 \leftrightarrow\phi_2$ accompanied by phase shifts.
This unitary transformation $b$ can be represented by the $2\times2$ block
\begin{equation}
b = \mmatrix{0}{e^{i(\alpha+\beta)}}{e^{i(\alpha-\beta)}}{0} = e^{i\alpha}\mmatrix{0}{e^{i\beta}}{e^{-i\beta}}{0}\,.
\end{equation}
Let us perform a basis change by writing $\phi_2 = e^{-i\beta} \tilde\phi_2$. 
Then, within the space of doublets $\tilde \phi_1 \equiv \phi_1$ and $\tilde\phi_2$,
the transformation $a$ is unchanged while the transformation $b$ becomes 
$b = e^{i\alpha} \mmatrix{0}{1}{1}{0}$.

A special case of this trick is when $\alpha = \pi$. 
Bringing the minus sign inside the matrix,
we can interpret it as $\beta = \pi$ instead of $\alpha=\pi$. This allows us to flip the sign of one the two doublets
and arrive at $b = \mmatrix{0}{1}{1}{0}$.

The same trick can be applied to the longer permutation cycles. For example, 
order-3 permutation accompanied by arbitrary phase shifts can be transformed, 
upon redefinition of the doublets, to $c = e^{i\alpha}\times$ the cyclic permutation.


\begin{thebibliography}{99}

\bibitem{CMS:2012qbp}
S.~Chatrchyan \textit{et al.} [CMS],
Phys. Lett. B \textbf{716}, 30-61 (2012)
doi:10.1016/j.physletb.2012.08.021
[arXiv:1207.7235 [hep-ex]].

\bibitem{ATLAS:2012yve}
G.~Aad \textit{et al.} [ATLAS],
Phys. Lett. B \textbf{716}, 1-29 (2012)
doi:10.1016/j.physletb.2012.08.020
[arXiv:1207.7214 [hep-ex]].

\bibitem{Lee:1973iz}
T.~D.~Lee,
Phys. Rev. D \textbf{8}, 1226-1239 (1973)
doi:10.1103/PhysRevD.8.1226

\bibitem{Weinberg:1976hu}
S.~Weinberg,
Phys. Rev. Lett. \textbf{37}, 657 (1976)
doi:10.1103/PhysRevLett.37.657

\bibitem{Branco:2011iw}
G.~C.~Branco, P.~M.~Ferreira, L.~Lavoura, M.~N.~Rebelo, M.~Sher and J.~P.~Silva,
Phys. Rept. \textbf{516}, 1-102 (2012)
doi:10.1016/j.physrep.2012.02.002
[arXiv:1106.0034 [hep-ph]].

\bibitem{Ivanov:2017dad}
I.~P.~Ivanov,
Prog. Part. Nucl. Phys. \textbf{95}, 160-208 (2017)
doi:10.1016/j.ppnp.2017.03.001
[arXiv:1702.03776 [hep-ph]].


\bibitem{Ishimori:2010au}
H.~Ishimori, T.~Kobayashi, H.~Ohki, Y.~Shimizu, H.~Okada and M.~Tanimoto,
Prog. Theor. Phys. Suppl. \textbf{183}, 1-163 (2010)
doi:10.1143/PTPS.183.1
[arXiv:1003.3552 [hep-th]].

\bibitem{Bjorken:1977vt}
J.~D.~Bjorken and S.~Weinberg,
Phys. Rev. Lett. \textbf{38}, 622 (1977)
doi:10.1103/PhysRevLett.38.622

\bibitem{Wyler:1979fe}
D.~Wyler,
Phys. Rev. D \textbf{19}, 3369 (1979)
doi:10.1103/PhysRevD.19.3369

\bibitem{Leurer:1992wg}
M.~Leurer, Y.~Nir and N.~Seiberg,
Nucl. Phys. B \textbf{398}, 319-342 (1993)
doi:10.1016/0550-3213(93)90112-3
[arXiv:hep-ph/9212278 [hep-ph]].

\bibitem{GonzalezFelipe:2014mcf}
R.~Gonz\'alez Felipe, I.~P.~Ivanov, C.~C.~Nishi, H.~Ser\^odio and J.~P.~Silva,
Eur. Phys. J. C \textbf{74}, no.7, 2953 (2014)
doi:10.1140/epjc/s10052-014-2953-9
[arXiv:1401.5807 [hep-ph]].

\bibitem{GonzalezFelipe:2013yhh}
R.~Gonzalez Felipe, H.~Serodio and J.~P.~Silva,
Phys. Rev. D \textbf{88}, no.1, 015015 (2013)
doi:10.1103/PhysRevD.88.015015
[arXiv:1304.3468 [hep-ph]].

\bibitem{GonzalezFelipe:2013xok}
R.~Gonz\'alez Felipe, H.~Ser\^odio and J.~P.~Silva,
Phys. Rev. D \textbf{87}, no.5, 055010 (2013)
doi:10.1103/PhysRevD.87.055010
[arXiv:1302.0861 [hep-ph]].

\bibitem{Bree:2023ojl}
I.~Bree, S.~Carrolo, J.~C.~Romao and J.~P.~Silva,
Eur. Phys. J. C \textbf{83}, no.4, 292 (2023)
doi:10.1140/epjc/s10052-023-11463-5
[arXiv:2301.04676 [hep-ph]].

\bibitem{Pakvasa:1978tx}
S.~Pakvasa and H.~Sugawara,
Phys. Lett. B \textbf{82}, 105-107 (1979)
doi:10.1016/0370-2693(79)90436-2

\bibitem{Ma:1979we}
E.~Ma,
Phys. Rev. D \textbf{20}, 2408 (1979)
doi:10.1103/PhysRevD.20.2408

\bibitem{Derman:1979nf}
E.~Derman and H.~S.~Tsao,
Phys. Rev. D \textbf{20}, 1207 (1979)
doi:10.1103/PhysRevD.20.1207

\bibitem{Christos:1984if}
G.~A.~Christos,
Austral. J. Phys. \textbf{38}, 23 (1985)
doi:10.1071/PH850023

\bibitem{Rajpoot:1988gw}
S.~Rajpoot,
Phys. Rev. D \textbf{40}, 873 (1989)
doi:10.1103/PhysRevD.40.873

\bibitem{Kobayashi:1985ek}
T.~Kobayashi,
Lett. Nuovo Cim. \textbf{44}, 417 (1985)
doi:10.1007/BF02746707

\bibitem{Albright:1979yc}
C.~H.~Albright, J.~Smith and S.~H.~H.~Tye,
Phys. Rev. D \textbf{21}, 711 (1980)
doi:10.1103/PhysRevD.21.711

\bibitem{Drees:1988fc}
M.~Drees,
Int. J. Mod. Phys. A \textbf{4}, 3635 (1989)
doi:10.1142/S0217751X89001448

\bibitem{Haber:1989xc}
H.~E.~Haber and Y.~Nir,
Nucl. Phys. B \textbf{335}, 363-394 (1990)
doi:10.1016/0550-3213(90)90499-4

\bibitem{Griest:1989ew}
K.~Griest and M.~Sher,
Phys. Rev. Lett. \textbf{64}, 135 (1990)
doi:10.1103/PhysRevLett.64.135

\bibitem{Griest:1990vh}
K.~Griest and M.~Sher,
Phys. Rev. D \textbf{42}, 3834-3849 (1990)
doi:10.1103/PhysRevD.42.3834

\bibitem{Masip:1995sm}
M.~Masip and A.~Rasin,
Phys. Rev. D \textbf{52}, R3768-R3772 (1995)
doi:10.1103/PhysRevD.52.R3768
[arXiv:hep-ph/9506471 [hep-ph]].

\bibitem{Grossman:1994jb}
Y.~Grossman,
Nucl. Phys. B \textbf{426}, 355-384 (1994)
doi:10.1016/0550-3213(94)90316-6
[arXiv:hep-ph/9401311 [hep-ph]].

\bibitem{Ma:1990qh}
E.~Ma,
Phys. Rev. D \textbf{43}, 2761-2764 (1991)
doi:10.1103/PhysRevD.43.R2761

\bibitem{Ma:1991eg}
E.~Ma,
Phys. Rev. D \textbf{44}, 587-589 (1991)
doi:10.1103/PhysRevD.44.R587

\bibitem{Duong:1991yt}
T.~V.~Duong and E.~Ma,
Phys. Rev. D \textbf{45}, 2570-2573 (1992)
doi:10.1103/PhysRevD.45.2570

\bibitem{Franklin:1995pk}
A.~Franklin,
Rev. Mod. Phys. \textbf{67}, 457-490 (1995)
doi:10.1103/RevModPhys.67.457

\bibitem{Deshpande:1991zh}
N.~G.~Deshpande, M.~Gupta and P.~B.~Pal,
Phys. Rev. D \textbf{45}, 953-957 (1992)
doi:10.1103/PhysRevD.45.953

\bibitem{Lavoura:1991yh}
L.~Lavoura,
Phys. Rev. D \textbf{44}, 1610-1612 (1991)
doi:10.1103/PhysRevD.44.1610

\bibitem{Lavoura:1992xj}
L.~Lavoura,
Phys. Rev. D \textbf{46}, 4101-4103 (1992)
doi:10.1103/PhysRevD.46.4101

\bibitem{Deshpande:1993py}
N.~G.~Deshpande and X.~G.~He,
Phys. Rev. D \textbf{49}, 4812-4819 (1994)
doi:10.1103/PhysRevD.49.4812
[arXiv:hep-ph/9312271 [hep-ph]].

\bibitem{Lavoura:1999dn}
L.~Lavoura,
Phys. Rev. D \textbf{61}, 077303 (2000)
doi:10.1103/PhysRevD.61.077303
[arXiv:hep-ph/9907538 [hep-ph]].

\bibitem{Cree:2011uy}
G.~Cree and H.~E.~Logan,
Phys. Rev. D \textbf{84}, 055021 (2011)
doi:10.1103/PhysRevD.84.055021
[arXiv:1106.4039 [hep-ph]].

\bibitem{Arroyo-Urena:2019lzv}
M.~A.~Arroyo-Ure\~na, J.~L.~Diaz-Cruz, B.~O.~Larios-L\'opez and M.~A.~P.~de Le\'on,
Chin. Phys. C \textbf{45}, no.2, 023118 (2021)
doi:10.1088/1674-1137/abcfae
[arXiv:1901.01304 [hep-ph]].

\bibitem{Rodejohann:2019izm}
W.~Rodejohann and U.~Salda\~na-Salazar,
JHEP \textbf{07}, 036 (2019)
doi:10.1007/JHEP07(2019)036
[arXiv:1903.00983 [hep-ph]].

\bibitem{Goncalves:2023ydf}
B.~L.~Gon\c{c}alves, M.~Knauss and M.~Sher,
[arXiv:2301.08641 [hep-ph]].

\bibitem{Porto:2007ed}
R.~A.~Porto and A.~Zee,
Phys. Lett. B \textbf{666}, 491-495 (2008)
doi:10.1016/j.physletb.2008.08.001
[arXiv:0712.0448 [hep-ph]].

\bibitem{Porto:2008hb}
R.~A.~Porto and A.~Zee,
Phys. Rev. D \textbf{79}, 013003 (2009)
doi:10.1103/PhysRevD.79.013003
[arXiv:0807.0612 [hep-ph]].

\bibitem{CarcamoHernandez:2021osw}
A.~E.~C\'arcamo Hern\'andez, I.~de Medeiros Varzielas, M.~L.~L\'opez-Ib\'a\~nez and A.~Melis,
JHEP \textbf{05}, 215 (2021)
doi:10.1007/JHEP05(2021)215
[arXiv:2102.05658 [hep-ph]].

\bibitem{Vien:2020aif}
V.~V.~Vien,
Nucl. Phys. B \textbf{956}, 115015 (2020)
doi:10.1016/j.nuclphysb.2020.115015

\bibitem{Deshpande:1977rw}
N.~G.~Deshpande and E.~Ma,
Phys. Rev. D \textbf{18}, 2574 (1978)
doi:10.1103/PhysRevD.18.2574

\bibitem{Meloni:2011cc}
D.~Meloni, S.~Morisi and E.~Peinado,
Phys. Lett. B \textbf{703}, 281-287 (2011)
doi:10.1016/j.physletb.2011.07.084
[arXiv:1104.0178 [hep-ph]].

\bibitem{Lavoura:2011ry}
L.~Lavoura,
J. Phys. G \textbf{39}, 025202 (2012)
doi:10.1088/0954-3899/39/2/025202
[arXiv:1109.6854 [hep-ph]].

\bibitem{Meloni:2010sk}
D.~Meloni, S.~Morisi and E.~Peinado,
Phys. Lett. B \textbf{697}, 339-342 (2011)
doi:10.1016/j.physletb.2011.02.019
[arXiv:1011.1371 [hep-ph]].

\bibitem{Boucenna:2011tj}
M.~S.~Boucenna, M.~Hirsch, S.~Morisi, E.~Peinado, M.~Taoso and J.~W.~F.~Valle,
JHEP \textbf{05}, 037 (2011)
doi:10.1007/JHEP05(2011)037
[arXiv:1101.2874 [hep-ph]].

\bibitem{deAdelhartToorop:2011ad}
R.~de Adelhart Toorop, F.~Bazzocchi and S.~Morisi,
Nucl. Phys. B \textbf{856}, 670-681 (2012)
doi:10.1016/j.nuclphysb.2011.11.020
[arXiv:1104.5676 [hep-ph]].

\bibitem{Bonilla:2023pna}
C.~Bonilla, J.~Herms, O.~Medina and E.~Peinado,
[arXiv:2301.10811 [hep-ph]].

\bibitem{Ivanov:2012hc}
I.~P.~Ivanov and V.~Keus,
Phys. Rev. D \textbf{86}, 016004 (2012)
doi:10.1103/PhysRevD.86.016004
[arXiv:1203.3426 [hep-ph]].

\bibitem{Diaz-Cruz:2014pla}
J.~L.~Diaz-Cruz and U.~J.~Salda\~na-Salazar,
Nucl. Phys. B \textbf{913}, 942-963 (2016)
doi:10.1016/j.nuclphysb.2016.10.018
[arXiv:1405.0990 [hep-ph]].

\bibitem{Keus:2014isa}
V.~Keus, S.~F.~King and S.~Moretti,
Phys. Rev. D \textbf{90}, no.7, 075015 (2014)
doi:10.1103/PhysRevD.90.075015
[arXiv:1408.0796 [hep-ph]].

\bibitem{Ma:2001dn}
E.~Ma and G.~Rajasekaran,
Phys. Rev. D \textbf{64}, 113012 (2001)
doi:10.1103/PhysRevD.64.113012
[arXiv:hep-ph/0106291 [hep-ph]].

\bibitem{He:2006dk}
X.~G.~He, Y.~Y.~Keum and R.~R.~Volkas,
JHEP \textbf{04}, 039 (2006)
doi:10.1088/1126-6708/2006/04/039
[arXiv:hep-ph/0601001 [hep-ph]].

\bibitem{Grimus:2008tt}
W.~Grimus and L.~Lavoura,
JHEP \textbf{09}, 106 (2008)
doi:10.1088/1126-6708/2008/09/106
[arXiv:0809.0226 [hep-ph]].

\bibitem{Grimus:2008nf}
W.~Grimus and L.~Lavoura,
Phys. Lett. B \textbf{671}, 456-461 (2009)
doi:10.1016/j.physletb.2008.12.041
[arXiv:0810.4516 [hep-ph]].

\bibitem{Grimus:2008vg}
W.~Grimus and L.~Lavoura,
JHEP \textbf{04}, 013 (2009)
doi:10.1088/1126-6708/2009/04/013
[arXiv:0811.4766 [hep-ph]].

\bibitem{Grimus:2009sq}
W.~Grimus and L.~Lavoura,
Phys. Lett. B \textbf{687}, 188-193 (2010)
doi:10.1016/j.physletb.2010.03.025
[arXiv:0912.4361 [hep-ph]].

\bibitem{Grimus:2009pg}
W.~Grimus, L.~Lavoura and P.~O.~Ludl,
J. Phys. G \textbf{36}, 115007 (2009)
doi:10.1088/0954-3899/36/11/115007
[arXiv:0906.2689 [hep-ph]].

\bibitem{Ferreira:2011hw}
P.~M.~Ferreira and L.~Lavoura,
[arXiv:1111.5859 [hep-ph]].

\bibitem{Park:2011zt}
N.~W.~Park, K.~H.~Nam and K.~Siyeon,
Phys. Rev. D \textbf{83}, 056013 (2011)
doi:10.1103/PhysRevD.83.056013
[arXiv:1101.4134 [hep-ph]].

\bibitem{BenTov:2012tg}
Y.~BenTov, X.~G.~He and A.~Zee,
JHEP \textbf{12}, 093 (2012)
doi:10.1007/JHEP12(2012)093
[arXiv:1208.1062 [hep-ph]].

\bibitem{Grossman:2014xia}
Y.~Grossman and C.~Peset,
JHEP \textbf{04}, 033 (2014)
doi:10.1007/JHEP04(2014)033
[arXiv:1401.1818 [hep-ph]].

\bibitem{Nelson:1993vc}
A.~E.~Nelson and L.~Randall,
Phys. Lett. B \textbf{316}, 516-520 (1993)
doi:10.1016/0370-2693(93)91037-N
[arXiv:hep-ph/9308277 [hep-ph]].

\bibitem{Krasnikov:1993qd}
N.~Krasnikov, G.~Kreyerhoff and R.~Rodenberg,
Nuovo Cim. A \textbf{107}, 589-596 (1994)
doi:10.1007/BF02768793

\bibitem{Aranda:2000zf}
A.~Aranda and M.~Sher,
Phys. Rev. D \textbf{62}, 092002 (2000)
doi:10.1103/PhysRevD.62.092002
[arXiv:hep-ph/0005113 [hep-ph]].

\bibitem{Marshall:2010qi}
G.~Marshall and M.~Sher,
Phys. Rev. D \textbf{83}, 015005 (2011)
doi:10.1103/PhysRevD.83.015005
[arXiv:1011.3016 [hep-ph]].

\bibitem{Kawase:2011az}
H.~Kawase,
JHEP \textbf{12}, 094 (2011)
doi:10.1007/JHEP12(2011)094
[arXiv:1110.3861 [hep-ph]].

\bibitem{Clark:2011cv}
T.~E.~Clark, S.~T.~Love and T.~ter Veldhuis,
Phys. Rev. D \textbf{85}, 015014 (2012)
doi:10.1103/PhysRevD.85.015014
[arXiv:1107.3116 [hep-ph]].

\bibitem{Yagyu:2012qp}
K.~Yagyu,
[arXiv:1204.0424 [hep-ph]].

\bibitem{Dutta:2018yos}
B.~Dutta and Y.~Mimura,
Phys. Lett. B \textbf{790}, 589-594 (2019)
doi:10.1016/j.physletb.2019.01.065
[arXiv:1810.08413 [hep-ph]].

\bibitem{Chacko:2005pe}
Z.~Chacko, H.~S.~Goh and R.~Harnik,
Phys. Rev. Lett. \textbf{96}, 231802 (2006)
doi:10.1103/PhysRevLett.96.231802
[arXiv:hep-ph/0506256 [hep-ph]].

\bibitem{Chacko:2005vw}
Z.~Chacko, Y.~Nomura, M.~Papucci and G.~Perez,
JHEP \textbf{01}, 126 (2006)
doi:10.1088/1126-6708/2006/01/126
[arXiv:hep-ph/0510273 [hep-ph]].

\bibitem{Falkowski:2006qq}
A.~Falkowski, S.~Pokorski and M.~Schmaltz,
Phys. Rev. D \textbf{74}, 035003 (2006)
doi:10.1103/PhysRevD.74.035003
[arXiv:hep-ph/0604066 [hep-ph]].

\bibitem{Chang:2006ra}
S.~Chang, L.~J.~Hall and N.~Weiner,
Phys. Rev. D \textbf{75}, 035009 (2007)
doi:10.1103/PhysRevD.75.035009
[arXiv:hep-ph/0604076 [hep-ph]].

\bibitem{Yu:2016bku}
J.~H.~Yu,
Phys. Rev. D \textbf{94}, no.11, 111704 (2016)
doi:10.1103/PhysRevD.94.111704
[arXiv:1608.01314 [hep-ph]].

\bibitem{Yu:2016swa}
J.~H.~Yu,
JHEP \textbf{12}, 143 (2016)
doi:10.1007/JHEP12(2016)143
[arXiv:1608.05713 [hep-ph]].

\bibitem{Yu:2016cdr}
J.~H.~Yu,
Phys. Rev. D \textbf{95}, no.9, 095028 (2017)
doi:10.1103/PhysRevD.95.095028
[arXiv:1612.09300 [hep-ph]].

\bibitem{Belanger:2022esk}
G.~B\'elanger, A.~Pukhov, C.~E.~Yaguna and \'O.~Zapata,
JHEP \textbf{03}, 100 (2023)
doi:10.1007/JHEP03(2023)100
[arXiv:2212.07488 [hep-ph]].

\bibitem{Ivanov:2011ae}
I.~P.~Ivanov, V.~Keus and E.~Vdovin,
J. Phys. A \textbf{45}, 215201 (2012)
doi:10.1088/1751-8113/45/21/215201
[arXiv:1112.1660 [math-ph]].

\bibitem{Ivanov:2012fp}
I.~P.~Ivanov and E.~Vdovin,
Eur. Phys. J. C \textbf{73}, no.2, 2309 (2013)
doi:10.1140/epjc/s10052-013-2309-x
[arXiv:1210.6553 [hep-ph]].

\bibitem{Ivanov:2012ry}
I.~P.~Ivanov and E.~Vdovin,
Phys. Rev. D \textbf{86}, 095030 (2012)
doi:10.1103/PhysRevD.86.095030
[arXiv:1206.7108 [hep-ph]].

\bibitem{Ivanov:2014doa}
I.~P.~Ivanov and C.~C.~Nishi,
JHEP \textbf{01}, 021 (2015)
doi:10.1007/JHEP01(2015)021
[arXiv:1410.6139 [hep-ph]].

\bibitem{Darvishi:2019dbh}
N.~Darvishi and A.~Pilaftsis,
Phys. Rev. D \textbf{101}, no.9, 095008 (2020)
doi:10.1103/PhysRevD.101.095008
[arXiv:1912.00887 [hep-ph]].

\bibitem{Darvishi:2021txa}
N.~Darvishi, M.~R.~Masouminia and A.~Pilaftsis,
Phys. Rev. D \textbf{104}, no.11, 115017 (2021)
doi:10.1103/PhysRevD.104.115017
[arXiv:2106.03159 [hep-ph]].

\bibitem{Ivanov:2015mwl}
I.~P.~Ivanov and J.~P.~Silva,
Phys. Rev. D \textbf{93}, no.9, 095014 (2016)
doi:10.1103/PhysRevD.93.095014
[arXiv:1512.09276 [hep-ph]].

\bibitem{Haber:2018iwr}
H.~E.~Haber, O.~M.~Ogreid, P.~Osland and M.~N.~Rebelo,
JHEP \textbf{01}, 042 (2019)
doi:10.1007/JHEP01(2019)042
[arXiv:1808.08629 [hep-ph]].

\bibitem{deMedeirosVarzielas:2019rrp}
I.~de Medeiros Varzielas and I.~P.~Ivanov,
Phys. Rev. D \textbf{100}, no.1, 015008 (2019)
doi:10.1103/PhysRevD.100.015008
[arXiv:1903.11110 [hep-ph]].

\bibitem{Ecker:1981wv}
G.~Ecker, W.~Grimus and W.~Konetschny,
Nucl. Phys. B \textbf{191}, 465-492 (1981)
doi:10.1016/0550-3213(81)90309-6

\bibitem{Ecker:1983hz}
G.~Ecker, W.~Grimus and H.~Neufeld,
Nucl. Phys. B \textbf{247}, 70-82 (1984)
doi:10.1016/0550-3213(84)90373-0

\bibitem{Ecker:1987qp}
G.~Ecker, W.~Grimus and H.~Neufeld,
J. Phys. A \textbf{20}, L807 (1987)
doi:10.1088/0305-4470/20/12/010

\bibitem{Grimus:1995zi}
W.~Grimus and M.~N.~Rebelo,
Phys. Rept. \textbf{281}, 239-308 (1997)
doi:10.1016/S0370-1573(96)00030-0
[arXiv:hep-ph/9506272 [hep-ph]].

\bibitem{Branco:1999fs}
G.~C.~Branco, L.~Lavoura and J.~P.~Silva,
Int. Ser. Monogr. Phys. \textbf{103}, 1-536 (1999)

\bibitem{Ivanov:2018qni}
I.~P.~Ivanov and M.~Laletin,
Phys. Rev. D \textbf{98}, no.1, 015021 (2018)
doi:10.1103/PhysRevD.98.015021
[arXiv:1804.03083 [hep-ph]].

\bibitem{Isaacs}
I.~Isaacs, Finite Group Theory (Am. Math. Soc., Providence,
2008).

\bibitem{GAP}
  The GAP Group.
  \textit{GAP---Groups, Algorithms, Programming---A System
    for Computational Discrete Algebra. Version 4.11.1; 2022}.
  Available at 
 {\tt https://www.gap-system.org}.

\bibitem{SmallGroups}
  H.~U.~Besche, B.~Eick and E.~O'Brien,
  \textit{The SmallGroups Library. Version 1.5.1; 2022}.
  Available at {\tt https://www.gap-system.org/Packages/smallgrp.html}.

\bibitem{Luhn:2007sy}
C.~Luhn, S.~Nasri and P.~Ramond,
Phys. Lett. B \textbf{652}, 27-33 (2007)
doi:10.1016/j.physletb.2007.06.059
[arXiv:0706.2341 [hep-ph]].

\bibitem{Hagedorn:2008bc}
C.~Hagedorn, M.~A.~Schmidt and A.~Y.~Smirnov,
Phys. Rev. D \textbf{79}, 036002 (2009)
doi:10.1103/PhysRevD.79.036002
[arXiv:0811.2955 [hep-ph]].

\bibitem{Cao:2010mp}
Q.~H.~Cao, S.~Khalil, E.~Ma and H.~Okada,
Phys. Rev. Lett. \textbf{106}, 131801 (2011)
doi:10.1103/PhysRevLett.106.131801
[arXiv:1009.5415 [hep-ph]].

\bibitem{Vien:2014gza}
V.~V.~Vien and H.~N.~Long,
JHEP \textbf{04}, 133 (2014)
doi:10.1007/JHEP04(2014)133
[arXiv:1402.1256 [hep-ph]].

\bibitem{Bonilla:2014xla}
C.~Bonilla, S.~Morisi, E.~Peinado and J.~W.~F.~Valle,
Phys. Lett. B \textbf{742}, 99-106 (2015)
doi:10.1016/j.physletb.2015.01.017
[arXiv:1411.4883 [hep-ph]].

\bibitem{Ivanov:2013bka}
I.~P.~Ivanov and C.~C.~Nishi,
JHEP \textbf{11}, 069 (2013)
doi:10.1007/JHEP11(2013)069
[arXiv:1309.3682 [hep-ph]].

\bibitem{TheCode}
Jiazhen Shao, The 4HDM Toolbox,
available at {\tt  https://github.com/JiazhenShao/4HDM-Toolbox.git}


\end{thebibliography}
\end{document}